\documentclass[12pt]{article}

\usepackage[margin=1in, paperwidth=10in, paperheight=11in]{geometry}
\usepackage[sc]{mathpazo}
\usepackage[T1]{fontenc}
\usepackage{amsfonts}
\usepackage{amsmath}
\usepackage{amssymb}
\usepackage{mathtools} 
\usepackage[titletoc]{appendix} 
\usepackage{fancyhdr} 
\usepackage[final]{graphicx}
\usepackage{ragged2e} 
\usepackage{enumitem} 
\usepackage{caption}
\usepackage{subcaption} 
\usepackage{booktabs} 
\usepackage{authblk} 
\usepackage{float}

\usepackage{natbib}
\setcitestyle{authoryear, open={(},close={)}} 
\usepackage{comment}

\usepackage[dvipsnames]{xcolor}
\definecolor{carnelian}{rgb}{0.0, 0.0, 1.0}
\usepackage[pagebackref=true]{hyperref}
\hypersetup{
	colorlinks=true,
	linkcolor=carnelian,
	filecolor=blue,
	citecolor = carnelian,      
	urlcolor=cyan,
}

\title{Monetary Policy, Digital Assets, and DeFi Activity}
\date{\Large{February 2023}}
\author[1]{Antzelos Kyriazis}
\author[2]{Iason Ofeidis}
\author[2]{Georgios Palaiokrassas}
\author[2]{Leandros Tassiulas}
\affil[1]{Department of Economics, Yale University}
\affil[2]{Department of Electrical Engineering and Yale Institute for Network Science, Yale University}
 
\begin{document}
\maketitle
 
\thispagestyle{empty}

\begin{abstract}
     This paper studies the effects of unexpected changes in US monetary policy on digital asset returns. We use event study regressions and find that monetary policy surprises negatively affect BTC and ETH, the two largest digital assets, but do not significantly affect the rest of the market. Second, we use high-frequency price data to examine the effect of the FOMC statements release and Minutes release on the prices of the assets with the higher collateral usage on the Ethereum Blockchain Decentralized Finance (DeFi) ecosystem. The FOMC statement release strongly affects the volatility of digital asset returns, while the effect of the Minutes release is weaker. The volatility effect strengthened after December 2021, when the Federal Reserve changed its policy to fight inflation. We also show that some borrowing interest rates in the Ethereum DeFi ecosystem are affected positively by unexpected changes in monetary policy. In contrast, the debt outstanding and the total value locked are negatively affected. Finally, we utilize a local Ethereum Blockchain node to record the activity history of primary DeFi functions, such as depositing, borrowing, and liquidating, and study how these are influenced by the FOMC announcements over time.

\end{abstract}

\newpage

\section{Introduction}

Digital assets constitute a relatively new asset class that emerged after the creation of the Bitcoin network in 2009. During 2009-2019, this new asset class was considered uncorrelated with the traditional asset classes. There was a general belief that macroeconomic conditions and policies did not significantly affect the prices and the returns on these assets. However, since late 2021 the movements of digital asset prices have been following step-by-step the downward movements of the US stock market, which can be primarily attributed to the higher interest rates announced by the Federal Reserve in response to high inflation. The effects of the FOMC announcements on traditional asset prices and returns have been thoroughly studied. However, it is still unknown what are the effects of the FOMC announcements on digital asset prices and returns. 

This paper provides insights into the above question. First, following an event study approach over the FOMC announcement dates for the period 01/01/2018 - 12/21/2022, we examine the effects of the total and the unexpected changes in the federal funds rate on the returns of BTC and ETH, the two largest digital assets by market cap, and also the effects that these changes in the federal funds rate have on the returns of the whole market and the market without the two largest coins. We also examine how the NASDAQ index is affected by the changes in monetary policy in order to compare with the effects on digital assets, given the very high correlation we find between digital asset prices and the NASDAQ index. We find that unexpected increases in the federal funds rate have negative and statistically significant effects on the returns to BTC and ETH over the period examined. These negative effects have a slightly higher magnitude than the negative effects caused by monetary policy surprises on the returns to the NASDAQ index. In our view, this is an indication that at least the two largest coins have been priced and have behaved as risky technology stocks with respect to the announcements of the Federal Reserve, although their network nature calls for different valuation methods that would take into account the value of the network as well.

For the returns of the digital asset market as a whole and the returns of the digital asset market without BTC and ETH, we find that surprises in monetary policy have negative effects of similar magnitude as in the cases of BTC, ETH, and NASDAQ returns. However, now these effects are not statistically significant, although the $t$-statistics are close to values that would make these effects statistically significant. Of course, our analysis is based on the orthogonality assumption, which implies that monetary policy does not respond simultaneously to changes in the prices of digital assets. We believe that this is true, given the mandate of the Federal Reserve on achieving price stability and full employment but not asset price stability, and also given the relatively small size of the digital assets market, which does not create systemic risks that could justify interventions from the Federal Reserve.

The second method we use to examine how the announcements of the Federal Reserve affect digital asset returns is high-frequency volatility analysis. Specifically, we gather high-frequency data on the prices of the digital assets with the higher collateral use in the DeFi ecosystem of the Ethereum network and compute the 5-minute returns over an interval that starts two hours before the FOMC statement release or the Minutes release and ends two hours after the corresponding release. Then we compute the volatility of these 5-minute returns for each 5-minute interval over the release dates and compare it to the volatility in non-release dates, which are defined as the same days of the week, one week before and one week after the releases. We find that the FOMC statement release significantly increases the volatility of digital asset returns, especially at the time of the release and the time of the press conference. The increase is more significant for altcoins due to their riskier nature. The Minutes release, on the other hand, also increases the volatility of digital asset returns but less significantly than the FOMC statement release. We also find that for both types of releases, the volatility effect became stronger after 12/15/2021, when the Federal Reserve signaled a change to its low-interest rates policy of the previous years to fight inflation. 

We inspect how the unexpected changes in the federal funds rate affect the interest rates on DeFi platforms, the debt outstanding, and the total value locked (TVL). An event study for regressions related to the borrowing rates in the Compound
protocol is conducted, which shows that two out of five of the assets studied relate in a statistically significant way to the unexpected changes in federal funds rate. Furthermore, we perform regression analysis for the natural logarithm for both the total debt outstanding and TVL for a DeFi platform, which results in a statistically significant relationship for the second one.

Finally, we investigate how the FOMC announcements influence the activity in DeFi. We begin by selecting four of the leading platforms in the DeFi scene (Aave, Compound, Maker and Liquity) based on their Total Value Locked. We focus on some of their fundamental actions, such as depositing and withdrawing funds, borrowing and liquidating. As a next step, we query these protocols for all of these actions that happened since January 2021 and we record their number of occurrences. Then, we plot these values in time versus the dates of FOMC announcements. From these graphs, we can observe how some actions are impacted by these announcements and how these unexpected changes in monetary policy seem to establish trends in DeFi activity in 2022.

\justify \textbf{Related Literature:} This paper is related to many different parts of the literature. First, it is related to papers that focus on the effects of unexpected monetary policy announcements on asset prices, such as \cite{kuttner2001monetary}, \cite{demiralp2004response}, \cite{bernanke2005explains} and \cite{gurkaynak2004actions}. Our work differs in that we are interested in how unexpected changes in monetary policy affect the prices and returns of digital assets, as well as other variables in the DeFi ecosystem of the Ethereum network, such as the total value locked, the total outstanding debt, and borrowing interest rates. Also, for this work, we do not focus on the whole path of monetary policy to explain the movements in digital asset prices as \cite{gurkaynak2004actions} do. The paper is also related to papers that use high-frequency data to analyze the effects of monetary policy announcements on asset prices and interest rates, such as \cite{fleming2005monetary} and \cite{rosa2013financial}. Again our work differs from these previous papers in that we focus on digital asset prices and interest rates in the DeFi ecosystem on the Ethereum network. 

Another line of related studies consists of papers such as \cite{aboura2022note}, in which the author uses linear regression models to explore the spillover effects between the levels of the federal funds rate and BTC returns. We differ from this study since we are using the unexpected changes in the federal funds rate on FOMC announcement days to explore the reaction of digital asset returns. \cite{pyo2020fomc}, on the  other hand, use a set of dummy variables representing the corresponding days before, of, and after the FOMC announcement to examine the announcement effects on BTC. We differ by using the unexpected changes in the federal funds rate, and also by examining the effects on the returns on ETH, as well as on market indices.

\justify \textbf{Related Work about DeFi:}
While there has been an extended research literature on blockchain, DeFi is a more recent research area with fewer works focusing on different aspects of this ecosystem. Research efforts applied algorithms and analyzed the DeFi protocols and publicly available transaction data from different perspectives.
\cite{qin_empirical_2021} studied the breadth of the borrowing and lending markets of the Ethereum DeFi ecosystem, focusing on Aave, Compound, MakerDAO, and dYdX protocols, which collectively represented at that point over 85\% of the lending market on Ethereum. \cite{darlin2022debt} proposed an Ethereum address grouping algorithm and a classification algorithm based on DeFi protocols to calculate the percentage of fund flows into DeFi lending platforms that can be attributed to debt created elsewhere in the system ("debt-financed collateral"). Based on their system-level aggregate analysis from four major DeFi protocols' transactions (Aave, Compound, MakerDAO and Uniswap) they concluded that the widespread use of stablecoins as debt-financed collateral increases financial stability risks in the DeFi ecosystem. 

Other efforts analyzed front-running, back-running or sandwiching in blockchain systems and the practice of exploiting information that may change the price of an asset for financial gain such as \cite{zhou2021just}. Like high-frequency traders on Wall Street, arbitrage bots in blockchain systems, specifically in decentralized exchanges exploit inefficiencies, paying high transaction fees and optimizing network latency to front-run, i.e., anticipate and exploit, ordinary users’ trades. \cite{daian2020flash}, introduced the concept of gas price auctions among trading bots as well as the concept of miner extractable value (MEV). On another direction, the prediction of cryptocurrency prices and returns has been studied by applying different machine learning techniques and analyzing available frequencies and data such as \cite{akyildirim2021prediction}.

\justify \textbf{Contribution:} Overall our paper makes the following contributions:
\begin{enumerate}
    \item We provide event-study regressions based on the FOMC announcement days to estimate the effects of unexpected changes in monetary policy on the returns to BTC, ETH, the whole digital assets market, and the market without BTC and ETH.
    \item We use high-frequency volatility analysis to show how the FOMC statement release and the Minutes release affect the volatility of digital asset returns.
    \item We provide new insights on liquidation events by examining the characteristics of liquidated investors as well as the characteristics of their liquidators.
    \item We examine how the unexpected changes in the federal funds rate affect the DeFi borrowing rates, the outstanding debt, and the TVL in the Ethereum network.
    \item We investigate how the FOMC announcements influence the activity in DeFi.
\end{enumerate}

\justify \textbf{Paper Organization:} The rest of the paper is organized in the following way. Section 2 provides some background on blockchain technology, the Ethereum network, the decentralized applications built on it, and decentralized finance. Section 3 contains the event study regressions for BTC, ETH, and the market indices. Section 4 includes the high-frequency volatility analysis for various digital assets. Section 5 shows how decentralized finance activity evolves around the FOMC announcements and how unexpected changes in monetary policy affect interest rates, outstanding debt, and TVL in the Ethereum network. Section 6 concludes.

\section{Background}
Before we examine the effects of monetary policy announcements on digital asset returns, we provide some background on the blockchain technology, the two largest network protocols, Bitcoin and Ethereum, the decentralized applications, and decentralized finance.

\subsection{Blockchain and Bitcoin}
Blockchain was originally proposed by \cite{nakamoto2008bitcoin}, as the accounting method for Bitcoin cryptocurrency in 2008  and after its quiet launch, it grew to comprise billions of dollars of economic value (\cite{bonneau2015sok}). In the years that followed more blockchains and cryptocurrencies or 'altcoins' were introduced. There are two main categories of blockchains: i) permissionless like Bitcoin and Ethereum, where any user is able to join without permission; ii) permissioned which are not publicly accessible or have an access control layer.

Blockchain is a distributed ledger or distributed database recording digital transactions between two parties without the need for Third Trusted Parties (TTP). This way users interact with each other directly, without an intermediary (e.g. administrator, bank), while anonymity is preserved as a key feature. A cryptocurrency could be regarded as a digital asset or a medium of exchange using cryptography for preventing tampering. 

Blockchain is actually a hash-linked chain of blocks operating over a peer-to-peer (P2P) network. A block is a data structure and contains a timestamp, the transaction data and the hash of the previous block. Users willing to transfer assets broadcast digitally signed transactions, which are collected by the miners, formed into blocks, and appended to the blockchain. During this process, different rules are enforced by the protocols and the whole network follows a distributed consensus protocol. In Bitcoin, Proof of Work is achieved, where miners are competing to solve a computationally intensive puzzle, while in 2022 Ethereum blockchain transitioned from proof-of-work (PoW) to proof-of-stake (PoS) which is much less computationally intensive.

\subsection{Ethereum and Decentralized Applications}
In recent years, technology and ideas behind blockchain evolved and its potential for different applications beyond cryptocurrency became apparent such as for IoT application as studied by \cite{papadodimas2018implementation}, healthcare as studied by \cite{mcghin2019blockchain}, identity management by \cite{kuperberg2019blockchain}, telecommunications by \cite{chaer2019blockchain}, decentralized finance by \cite{perez2021liquidations} and many more applications as shown in \cite{hewa2021survey}.

A milestone for the course of blockchain technology was the development of the Ethereum project, offering new solutions by enabling smart contracts’ implementation and execution (\cite{wood2014ethereum}). It is an open-source decentralized platform that allows anyone to build and run decentralized applications that run on the nodes of the Ethereum network. It provides a virtual computing environment called Ethereum Virtual Machine but also a Turing complete programming language to write smart contracts. Smart Contracts are mainly written in the programming language solidity, while their logs and emitted events can be parsed and analyzed, as explained in detail in the section \ref{Data and Methodology}.

 In a currency perspective, Ethereum declares its native token called “Ether”, while many Tokens have also been introduced and run as ERC20-compliant smart contracts with the phenomenon of startups launching Initial Coin Offerings (ICOs) to raise funds, which led to hundreds of virtual tokens being distributed and traded on blockchains and exchanges, as shown in  \cite{victor2019measuring}.

\subsection{Decentralized Finance}
Ethereum smart contracts allow not only the creation of Decentralized Applications and Tokens but further the construction of sophisticated on-chain financial systems, namely Decentralized Finance (DeFi). In the DeFi ecosystem, any entity can deploy a financial protocol, by implementing the respective smart contracts and deploying it on the Ethereum blockchain network (\cite{qin_empirical_2021}). Users can interact with lending pools such as Aave (\cite{aave2023online}) and Maker (\cite{makerdao2023online}), Automated Market Maker exchanges such as Uniswap (\cite{Uniswap2023online}), stablecoins (\cite{usdc2023online}), derivatives, and asset management platforms. In DeFi debt is a popular form of leverage and while financial speculators often seek to increase their potential gains, debts entail the risks of liquidation, the process of selling the debt collateral at a discount to liquidators. 
Importing off-chain data into the blockchain virtual machine is of vital importance so that they are readable from smart contracts. To this direction, the Oracles have been introduced as a mechanism for including various types of data such as off-chain asset prices, e.g ETH/USD, as well as off-chain information needed to verify outcomes of prediction markets, as studied by \cite{werner2021sok}.

\section{Digital Asset Returns Reaction to Unexpected Changes in the Federal Funds Rate}

This section focuses on the impact of changes in monetary policy on digital asset prices, both for individual assets such as Bitcoin's BTC and Ethereum's ETH, which are the two assets with the highest institutional adoption\footnote{Bitcoin's BTC is an asset held in the balance sheet of various \href{https://buybitcoinworldwide.com/treasuries/}{public companies} including Tesla, Inc., Microstrategy, Coinbase Global, Inc., etc. It is also a \href{https://coinmarketcap.com/legal-tender-countries/}{legal tender} for countries such as El Salvador and the Central African Republic and an officially recognized means of payment in the canton of \href{https://planb.lugano.ch/accept-crypto-payments/}{Lugano} in Switzerland, and soon to be in \href{https://www.nasdaq.com/articles/brazil-approves-bill-regulating-use-of-bitcoin-as-payment}{Brazil}. Ethereum's ETH, together with BTC, are assets that are also offered by private investment institutions such as \href{https://www.fidelity.com/crypto/overview}{Fidelity} to their clients. \href{https://www.cnbc.com/2022/08/11/blackrock-launches-a-private-trust-to-give-clients-exposure-to-spot-bitcoin.html}{BlackRock} has also established a Bitcoin trust for its clients.} across the digital asset space and also for broad market indices. Of course, the prices of digital assets can change due to other events unrelated to monetary policy changes, such as an exploitation of a specific protocol which can create panic for the whole ecosystem related to this protocol. However, we are interested in the effect of unexpected monetary policy decisions on digital asset prices.

\begin{table}[t!]\centering
\def\sym#1{\ifmmode^{#1}\else\(^{#1}\)\fi}
\caption{Correlation Coefficients 01/01/2018 - 12/21/2022\label{correlation_coefficients}}
\begin{tabular}{l*{1}{ccccc}}
\toprule
                    &        NASDAQ&    NYXBT&      SPETH&         SPCBDM&         SPCBXM \\
\midrule
NASDAQ            &        1&   &    &     &       \\
\\
NYXBT               &        0.885&       1&     &   & \\
\\
SPETH             &        0.856&    0.925   &     1&   &  \\
\\
SPCBDM           &        0.865&       0.986&    0.965  &    1&  \\
\\
SPCBXM           &        0.639&       0.841&      0.883&    0.907   & 1\\

\bottomrule
\end{tabular}
\end{table}

\subsection{Event Study Results}

Our approach is an event study regressions approach following \cite{bernanke2005explains}. First, we gather data on the changes in the federal funds rate for the period 01/01/2018 to 12/21/2022 by using the statement releases\footnote{The reason why we decided to start the sample in 01/01/2018 and not in 01/01/2017 is because we believe that the real adoption of digital assets started after the burst of the speculative bubble of 2017 that made digital assets popular, and led to the bear market that gave the time and the focus to build the first decentralized finance applications and the first NFTs.} in the "Meeting Calendars, statements,  and minutes" website of the Federal Reserve.\footnote{The website can be accessed at \href{https://www.federalreserve.gov/monetarypolicy/fomccalendars.htm}{federalreserve.gov/monetarypolicy/fomccalendars.htm}} We focus on the FOMC statement releases that contain information about the federal funds rate and asset purchases, so those releases that provide information only about notation votes on issues other than the federal funds rate, and central bank asset purchases, are excluded. We decompose these changes into expected and unexpected changes. Our definition for unexpected changes is the same as in \cite{bernanke2005explains}. Specifically, we use the change in the current month's futures contract price scaled by a factor related to the days of the month so that

\begin{equation}
    \Delta i^u = \frac{D}{D-d}\left(f^0_{m,d}-f^0_{m,d-1}\right),
\end{equation}

\justify where in the previous equation $\Delta i^u$ is the unexpected target rate change, $f^0_{m,d}$ is the current-month futures rate, $D$ is the number of days in the month, $d$ is the specific day of interest, and $m$ is the specific month of interest. The expected change in the federal funds rate is then given by the difference between the total change and the unexpected change. 

\begin{table}[t!]\centering
\def\sym#1{\ifmmode^{#1}\else\(^{#1}\)\fi}
\caption{Summary Statistics - FOMC Announcement Days \label{summary}}
\begin{tabular}{l*{1}{ccccc}}
\toprule
                    &        Mean&          SD&         Min&         Max & Observations\\
\midrule
NYXBT Return \%           &        0.322&       4.241&     -15.462&       7.914&      43\\
\\
SPETH Return  \%         &        1.407&       4.814&     -13.462&      16.298&      43 \\
\\
SPCBXM Return \%           &        1.886&       4.203&      -7.267&      10.579&      43 \\
\\
SPCBDM Return \%           &        1.699&       3.825&      -7.758&      12.200&      43 \\
\\
NASDAQ Return  \%          &        0.0365&       2.562&     -12.321&       4.064&      43 \\
\\
Change in FFR    \%               &        0.0702&       0.310&      -0.850&       0.750&  43\\
\\
Expected Change in FFR     \%             &        0.0556&       0.340&      -1.063&       0.785&      43 \\
\\
Unexpected Change in FFR   \%              &         0.0146&       0.0576&      -0.0409&       0.299&      43  \\
\bottomrule
\end{tabular}
\end{table}

We also gather data on the changes in indices following the prices of BTC, ETH, and indices that track the whole market of digital assets and the market without BTC and ETH. Specifically, for the returns on BTC, we choose to use the NYSE Bitcoin Index (NYXBT) because this index has been tracked since 2015 by the New York Stock Exchange. For the returns on ETH, we could not find an index measured by the NYSE or an index that has been tracked since 2015, so we used the S\&P Ethereum Index (SPETH). For the market as a whole, we use the S\&P Cryptocurrency Broad Digital Market Index (SPCBDM). Finally, for the rest of the market without BTC and ETH, we use the S\&P Cryptocurrency BDM Ex-MegaCap Index (SPCBXM). We are interested in OLS estimation of linear models in which the dependent variable is the return on each of the previous indices on the announcement days, and the regressor is the total change in the federal funds rate

\begin{equation}
    r_t^j = \alpha_0 + \alpha_1\Delta i_t +\varepsilon_t,
\end{equation}

\justify or the unexpected change in the federal funds rate at this specific day, as in \cite{fleming2005monetary} and also in \cite{gurkaynak2004actions}, so that

\begin{equation}
    r_t^j = \beta_0 + \beta_1\Delta i_t^u +u_t.
\end{equation}

\justify where the superscript $j$ refers to any crypto index from those mentioned above. We also measure the correlation between the NASDAQ index and the aforementioned digital asset indices. We find a significant positive correlation between the movements of technology stocks and digital assets, as reported in Table \ref{correlation_coefficients}. Therefore, we run the same regressions using the NASDAQ index to compare the results for digital assets with those for technology stocks. The data for the NYXBT index, the NASDAQ index, and the current month futures rates were collected from Refinitiv. The rest indices were extracted from the S\&P Global website. 

Starting from Table \ref{summary}, which contains summary statistics for the indices of interest on FOMC announcement days, the average return on individual digital asset indices, digital asset market indices, and across asset classes is positive. The returns and the standard deviation of returns on digital assets are significantly higher than the respective NASDAQ measures. Market capitalization is one of the main reasons behind this result. The digital assets market is relatively small, and any significant positive or negative investments can move the market more relative to the stock market. 

Regarding the largest negative return across digital assets and asset classes on FOMC announcement days, this occured on 03/15/2020, when the COVID-19 shock hit the global financial markets. On the other hand, the maximum returns across digital assets and asset classes did not occur on the same days since the maximum return on ETH and the NASDAQ took place on 07/27/2022. In contrast, the other indices saw their maximum returns on 04/29/2020.

When it comes to monetary policy decisions, the average change in the federal funds rate over the period of interest is around 7 basis points with a standard deviation of 31 basis points. The largest increase is 75 basis points, a change that has occurred multiple times during 2022. The largest negative change is 85 basis points and took place on 03/15/2020. The expected change closely follows the total change in FFR in all categories, although they are not the same because of the unexpected part of the changes in the FFR.

\begin{table}[t!]\centering
\def\sym#1{\ifmmode^{#1}\else\(^{#1}\)\fi}
\caption{Regression Results - BTC and ETH\label{btc_regression}}
\begin{tabular*}{0.85\hsize}{@{}l @{\extracolsep{\fill}} *{4}{c}}
\toprule
                    &\multicolumn{1}{c}{(1)}&\multicolumn{1}{c}{(2)}&\multicolumn{1}{c}{(3)}&\multicolumn{1}{c}{(4)}\\
                    &\multicolumn{1}{c}{$ r^{\text{NYXBT}}$}&\multicolumn{1}{c}{$r^{\text{NYXBT}}$}&\multicolumn{1}{c}{$r^{\text{SPETH}}$}&\multicolumn{1}{c}{$r^{\text{SPETH}}$}\\
\midrule
$\Delta i$                   &       4.331\sym{*}  &                     &       5.416\sym{*}  &                     \\
                    &      (2.03)         &                     &      (2.38)         &                     \\
\addlinespace
$\Delta i^u$                  &                     &      -26.325\sym{*}  &                     &      -30.283\sym{*}  \\
                    &                     &     (-2.45)         &                     &     (-2.49)         \\
\addlinespace
Intercept            &      0.0175         &       0.707         &       1.026         &       1.850\sym{*}  \\
                    &      (0.03)         &      (1.12)         &      (1.44)         &      (2.54)         \\
\midrule
Adj. $R^2$         &          0.0782         &          0.1066         &          0.1002         &          0.1102         \\
\bottomrule
\multicolumn{5}{l}{\footnotesize \sym{*} \(p<0.05\), \sym{**} \(p<0.01\), \sym{***} \(p<0.001\)}\\
\end{tabular*}
\end{table}

Table \ref{btc_regression} contains the regression results for BTC and ETH. The indices tracking the returns on these two assets are affected negatively by surprises in monetary policy announcements. Specifically, an unexpected increase by one full percentage point in the federal funds rate reduces the return on BTC by 26.33\% and the return on ETH by 30.28\%. These results are significant at the 5\% significance level. The adjusted $R^2$ is similar for the two assets and close to 11\%. Hence, although the unexpected actions of the Federal Reserve matter for the volatility of the returns on these assets, it is far from being the most critical factor, which is not unexpected given that this asset class is new. There is much noise coming from idiosyncratic shocks hitting many of these assets. The results change when we consider the total change in the federal funds rate as a regressor. The responses of BTC and ETH returns become positive but significantly lower in absolute terms and remain statistically significant.

Table \ref{market_regression} contains the regression results for the market indices. These results are similar qualitatively to the results for BTC and ETH, which is not unexpected given the dominance of these two assets in the digital asset space, and the high correlation of the other digital assets with the two largest. The quantitative differences are also slight. Specifically, again a 1\% unexpected increase in the federal funds rate leads to a fall in the return of the broad market index by 19.53\%. The return on the market index that excludes BTC and ETH falls by 21\%. However, these results are not statistically significant at the 5\% significance level, but the $t$-statistics take values very close to the threshold values that would make the results statistically significant. 

Interestingly, when comparing the previous effects with the effects of unexpected monetary policy changes on the NASDAQ returns,  we find a slightly higher negative response of 23.98\%, which is also another argument that digital assets behave as if they were high-risk technological stocks. Nonetheless, the effect on the NASDAQ return is statistically significant even at the 0.1\% significance level. In addition, the unexpected change in the monetary policy explains 27.35\% of the volatility in the returns of the NASDAQ index. In contrast, for the digital asset market indices, this fraction is significantly lower and equal to 6.42\% for the broad market index and 6.06\% for the market index that excludes BTC and ETH. When examining the effects of the total change on interest rates, then again, those are positive and lower in magnitude, even for the NASDAQ index.

\begin{table}[t!]\centering
\def\sym#1{\ifmmode^{#1}\else\(^{#1}\)\fi}
\caption{Regression Results - Market Indices\label{market_regression}}
\begin{tabular*}{0.85\hsize}{@{}l @{\extracolsep{\fill}} *{6}{c} @{}}
\toprule
                    &\multicolumn{1}{c}{(1)}&\multicolumn{1}{c}{(2)}&\multicolumn{1}{c}{(3)}&\multicolumn{1}{c}{(4)}&\multicolumn{1}{c}{(5)}&\multicolumn{1}{c}{(6)}\\
                    &\multicolumn{1}{c}{$r^{\text{SPCBDM}}$}&\multicolumn{1}{c}{$r^{\text{SPCBDM}}$}&\multicolumn{1}{c}{$r^{\text{SPCBXM}}$}&\multicolumn{1}{c}{$r^{\text{SPCBXM}}$}&\multicolumn{1}{c}{$r^{\text{NASDAQ}}$}&\multicolumn{1}{c}{$r^{\text{NASDAQ}}$}\\
\midrule
$\Delta i$                   &       3.229         &                     &       3.929\sym{*}  &                     &       3.682\sym{**} &                     \\
                    &      (1.74)         &                     &      (1.94)         &                     &      (3.19)         &                     \\
\addlinespace
$\Delta i^u$                  &                     &      -19.53         &                     &      -21.008         &                     &      -23.980\sym{***}\\
                    &                     &     (-1.97)         &                     &     (-1.93)         &                     &     (-4.10)         \\
\addlinespace
Intercept            &       1.472\sym{*}  &       1.984\sym{**} &       1.610\sym{*}  &       2.193\sym{**} &      -0.222         &       0.387         \\
                    &      (2.52)         &      (3.41)         &      (2.53)         &      (3.42)         &     (-0.61)         &      (1.13)         \\
\midrule
Adj. $R^2$         &          0.0457         &          0.0642        &          0.0616         &          0.0606         &          0.1789         &          0.2735         \\
\bottomrule
\multicolumn{7}{l}{\footnotesize \sym{*} \(p<0.05\), \sym{**} \(p<0.01\), \sym{***} \(p<0.001\)}\\
\end{tabular*}
\end{table}

\subsection{Orthogonality}

The results based on the event-study methodology of subsection 3.1 depend on the assumption that the error term is orthogonal to the changes in the federal funds rate. One case in which this assumption would be violated is the case where monetary policy also responds to the changes in the digital assets market. However, we believe that this is not true. First, the mandate of the Federal Reserve dictates that monetary policy responds to ensure price stability and full employment, so there is no room for reaction to changes in asset prices in general and not only in the prices of digital assets. Second, even if the Federal Reserve could care about asset prices, the digital assets market is relatively small, with its peak market cap being close to \$3tn in November 2021. In other words, the digital assets market is not yet significant enough to create systemic risks that would make the Federal Reserve consider interventional measures. 

Another case in which the orthogonality condition would fail, as noted in \cite{bernanke2005explains}, is if monetary policy and the market for digital assets respond jointly and contemporaneously to new information such as the release of employment reports or reports about the price level, at the date of the monetary policy announcements. For instance, if new data releases show a high inflation rate, then this could lead the  Federal Reserve to increase its policy rate and, at the same time, make investors want to sell their riskier assets, including digital assets, in favor of safer ones, thinking that higher inflation will lead to interest rate hikes. As regards employment reports, according to the Bureau of Labor and Statistics,  during our sample period, there is only one date in which an employment report was released on the same date as the FOMC announcement, on 10/4/2019. On this day, an extra FOMC meeting took place. Although there was an unexpected change in monetary policy during this day, as indicated by the changes in the futures contracts, we decided not to proceed with this case since there is only one available observation. On the other hand, for price level reports, we turned to the release dates of the PCE price index. None of the monetary policy announcement dates in our sample coincided with the release dates of the  PCE index, so we conclude that orthogonality is not violated by a potential joint and contemporaneous response of monetary policy and the market of digital assets to new information about economic activity and inflation.


\begin{figure}[t!]
	\centering
	\caption{Residuals vs. Fitted Values}
	\includegraphics[width=6.5in]{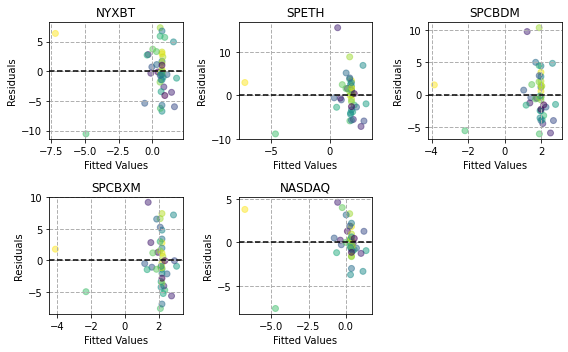} 
\end{figure}

\subsection{Heteroskedasticity}

The results presented in subsection 3.1 are based on simple regressions using the OLS estimator, assuming a constant variance for the error terms. It is well known that if this assumption is violated, then the OLS estimates remain unbiased, but the estimates of the errors, and as a result, the test statistics and the confidence intervals, can be biased, leading to wrong statistical inference. To see if heteroskedasticity is present in the error terms, we started by plotting the residuals vs. the fitted values for all the regressions discussed in the previous parts. 

Figure 1 contains the residuals vs. the fitted values for our previous regressions. Starting from the subgraphs for BTC and ETH, it is evident that, except for two outliers for the fitted values of the BTC return and three outliers for the fitted values of the ETH return, which, however, give values for the residuals close to the other fitted values, there is no clear pattern for the residuals since all of them are within a specific range independently of the fitted values. The results are very similar if we examine the market indices, including the NASDAQ index. Although from these graphs, one can conclude that there is no need for heteroskedasticity corrections, we also decided to run some White tests for heteroskedasticity. Table 5 contains the results.


According to Table \ref{tab5}, the White test predicts heteroskedasticity in the errors of the linear model for the NYXBT and the NASDAQ index. In contrast, there is no heteroskedasticity for the errors of the linear models for the SPETH index, the SPCBDM index, and the SPCBXM index. The White test predictions contrast with the graphs plotting the residuals vs. the fitted values. For instance, by comparing the two graphs for NYXBT and SPETH, we see that the behavior of the residuals is very similar, with more outliers for the SPETH index. However, the White test gives completely different results for the two indices. Moreover, the graphs of the residuals for the SPCBDM and the SPCBXM indices also look very similar, and the White test predicts no heteroskedasticity. Regarding the NASDAQ index, again, apart from the outliers, the graph looks very similar, but the White test predicts that there is heteroskedasticity. Overall, we believe that the White test rejects the homoskedasticity assumption for the NYXBT and the NASDAQ indices because of the one outlier residual in each case which is not within the range of values created by the rest residuals. Due to the influence of one observation in each case we are less inclined towards heteroskedasticity.

The White test prediction that there is heteroskedasticity in the residuals for the NYXBT regression would call either for a different estimator, such as the GLS, which requires knowing the form of heteroskedasticity to use, or for a correction in the standard errors using the robust standard errors. If we use robust standard errors, the estimated parameters do not change. However, the standard errors change, and the effects of the unexpected changes in monetary policy on the returns to BTC become statistically insignificant. Nonetheless, our sample consists of 43 observations, which might not be large enough for asymptotic analysis.

\begin{table}[t]\centering
\caption{Heteroskedasticity Test Results \label{tab5}}
\begin{tabular}{cccccc}
\toprule
Index & NYXBT & SPETH & SPCBDM & SPCBXM & NASDAQ \\ \hline 
$\chi^2$-statistic & $10.96$ & $2.01$ & $0.52$ & $0.28$ & $18.00$ \\
\\
$p$-value & $0.0042$ & $0.3661$ & $0.7710$ & $0.8676$ & $0.0001$ \\
\bottomrule
\end{tabular}
\end{table}

\subsection{Asymmetries}

In this subsection, we examine whether the response of digital asset prices and returns depends on the direction of the actions of the Federal Reserve. We are interested in the following three types of asymmetries: i) the response of digital asset prices depends on the sign of the surprise, ii) the response of digital asset prices depends on if the federal funds rate increases and iii) the response of digital asset prices depends on if the federal funds rate does not change. To test these three cases, we first create three dummy variables which take the value of 1 on the following three cases: i) the surprise in the federal funds rate change is positive,  ii) the change in the federal funds rate is positive, and iii) the change in the federal funds rate is zero. Then, we create three interactive variables defined as the product between the surprise element in the federal funds rate and each of the three dummy variables mentioned. We include these interactive terms, one at a time, in our regressions and report the results in Table \ref{btc_regression_asym} for BTC and ETH and Table \ref{market_regression_asym} for the market indices.

\begin{table}[t!]\centering
\def\sym#1{\ifmmode^{#1}\else\(^{#1}\)\fi}
\caption{Asymmetries Regression Results - BTC and ETH\label{btc_regression_asym}}
\begin{tabular*}{0.85\textwidth}{@{}l @{\extracolsep{\fill}} *{6}{c} @{}}
\toprule
                    &\multicolumn{1}{c}{(1)}&\multicolumn{1}{c}{(2)}&\multicolumn{1}{c}{(3)}&\multicolumn{1}{c}{(4)}&\multicolumn{1}{c}{(5)}&\multicolumn{1}{c}{(6)}\\
                    &\multicolumn{1}{c}{$ r^{\text{NYXBT}}$}&\multicolumn{1}{c}{$r^{\text{NYXBT}}$}&\multicolumn{1}{c}{$r^{\text{NYXBT}}$}&\multicolumn{1}{c}{$r^{\text{SPETH}}$}&\multicolumn{1}{c}{$r^{\text{SPETH}}$}&\multicolumn{1}{c}{$r^{\text{SPETH}}$}\\
\midrule
$\Delta i^u$                  &         20.12            &      -29.43\sym{*}  &  -27.72\sym{*}              &      50.69\sym{*} & -36.18\sym{**} & -32.20\sym{*}\\
                    &           (0.33)          &     (-2.70)         &     (-2.54)                &     (0.73)      & (-3.06)   & (-2.62)\\
\addlinespace
$\Delta i^u \times$ & & & & \\ 
\addlinespace
$\mathbb{1}_{\Delta i^u>0}$ &  -49.10 & & & -85.60 & & \\ 
 & (-0.77) & & & (-1.19) &  &  \\
\addlinespace
$\mathbb{1}_{\Delta i>0}$  & & 63.58 & & & 120.67\sym{**} & \\ 
  & & (1.33) & & & (2.32) \\
  \addlinespace
$\mathbb{1}_{\Delta i=0}$  & & & 70.82 & & & 97.15\\   
& & & (0.86)  & & & (1.04) \\
\addlinespace
Intercept            &      0.940         &       0.709         &       0.763         &       2.257\sym{**} & 1.855\sym{**} & 1.927\sym{*} \\
                    &      (1.34)         &      (1.13)         &      (1.20)         &      (2.86)   & (2.73)  & (2.68)    \\
\midrule
Adj. $R^2$         &          0.0975         &          0.1229         &          0.1007         &          0.1192     &  0.1960 &  0.1120 \\
\bottomrule
\multicolumn{5}{l}{\footnotesize \sym{*} \(p<0.05\), \sym{**} \(p<0.01\), \sym{***} \(p<0.001\)}\\
\end{tabular*}
\end{table}

According to Table \ref{btc_regression_asym}, including the interaction term corresponding to the case of positive surprises leads to non-significant regression results since now all the coefficients and even the intercept in the case of BTC have become statistically insignificant. Suppose we include the interaction term that considers positive changes in the federal funds rate. In that case, the results of the BTC return are similar, and the coefficient on the interaction term is not statistically significant. However, as regards the return on ETH, we observe an increase in absolute terms of the surprise element coefficient. In contrast, the coefficient of the interaction term is positive and statistically significant, taking a relatively large value of 120.67, implying that the direction of the change in the federal funds rate plays a significant role in the returns on ETH. In this regression, the adjusted $R^2$ is equal to 19.60\%, implying that a significant portion of the variation of the return on ETH can be explained by the unexpected movements in monetary policy, especially if these take place when the federal funds rate increases. Finally, as regards the interaction term corresponding to the case of zero change in the federal funds rate, we find that the results for the return on BTC are similar again, and the coefficient on the interaction term again is not statistically significant. This time the same is true for the return on ETH.

\begin{table}[t!]\centering
\def\sym#1{\ifmmode^{#1}\else\(^{#1}\)\fi}
\caption{Asymmetries Regression Results - Market Indices\label{market_regression_asym}}
\begin{tabular*}{\hsize}{@{}l @{\extracolsep{\fill}} *{9}{c} @{}}
\toprule
                    &\multicolumn{1}{c}{(1)}&\multicolumn{1}{c}{(2)}&\multicolumn{1}{c}{(3)}&\multicolumn{1}{c}{(4)}&\multicolumn{1}{c}{(5)}&\multicolumn{1}{c}{(6)}&\multicolumn{1}{c}{(7)}&\multicolumn{1}{c}{(8)}&\multicolumn{1}{c}{(9)}\\
                    &\multicolumn{1}{c}{$r^{\text{SPCBDM}}$}&\multicolumn{1}{c}{$r^{\text{SPCBDM}}$}&\multicolumn{1}{c}{$r^{\text{SPCBDM}}$}&\multicolumn{1}{c}{$r^{\text{SPCBXM}}$}&\multicolumn{1}{c}{$r^{\text{SPCBXM}}$}&\multicolumn{1}{c}{$r^{\text{SPCBXM}}$}&\multicolumn{1}{c}{$r^{\text{NASDAQ}}$}&\multicolumn{1}{c}{$r^{\text{NASDAQ}}$}&\multicolumn{1}{c}{$r^{\text{NASDAQ}}$}\\
\midrule
$\Delta i^u$                  &         50.77            &      -23.81\sym{*} & -20.82\sym{*}             &      -8.78 & -23.61\sym{*} & -22.08 & 9.73 & -26.74\sym{***} & -24.58\sym{***} \\
                    &           (0.90)          &     (-2.44)         &     (-2.07)                &     (-0.14)      & (-2.12)   & (-1.99) & (0.29) & (-4.68) & (-4.13)\\
\addlinespace
$\Delta i^u \times$   & & & & \\ 
\addlinespace
$\mathbb{1}_{\Delta i^u>0}$ &  -74.32 & & & -12.93 & & & -35.64 \\ 
 & (-1.27) & & & (-0.2) &  &  & (-1.03) \\
\addlinespace
$\mathbb{1}_{\Delta i>0}$  & & 87.66\sym{*} & & & 53.37\sym{**} & & & 56.53\sym{*}\\ 
  & & (2.04) & & & (1.09) & & & (2.25)\\
  \addlinespace
$\mathbb{1}_{\Delta i=0}$  & & & 65.51 & & & 54.19 & & & 30.52\\   
& & & (0.89)  & & & (0.64) & & & (0.67)\\
\addlinespace
Intercept            &      2.338\sym{**}         &       1.988\sym{**}         &       2.036\sym{**}         &       2.255\sym{**} & 2.195\sym{**} & 2.236\sym{**} & 0.557 & 0.390 & 0.412\\
                    &      (3.64)         &      (3.54)         &      (3.47)         &      (3.13)   & (3.43)  & (3.44)   & (1.46) & (1.19) & (1.18)\\
\midrule
Adj. $R^2$         &          0.0782         &          0.1312         &          0.0582         &          0.0380     &  0.0648 &  0.0469 & 0.2745 & 0.3391 &0.2638\\
\bottomrule
\multicolumn{5}{l}{\footnotesize \sym{*} \(p<0.05\), \sym{**} \(p<0.01\), \sym{***} \(p<0.001\)}\\
\end{tabular*}
\end{table}

Table \ref{market_regression_asym} contains the same exercises for the market indices. Again, the addition of the interaction term corresponding to positive surprises makes the results for the market indices statistically insignificant, and this is true even for the NASDAQ index. Suppose, instead, we include the interaction term corresponding to positive changes in the federal funds rate. In that case, the negative effect of the surprises in monetary policy on all market indices becomes stronger, as was the case with the returns on BTC and ETH. However, the coefficients on the surprise term of monetary policy are now statistically significant for the digital asset indices. The coefficients on the interaction term are positive and statistically significant for the total market indices but not for the crypto index that excludes BTC and ETH, implying again that an increase in the federal funds rate, when unexpected, can affect the markets significantly. Finally, as regards the interaction term corresponding to zero changes in the federal funds rate, this term is not statistically significant in any of the regressions related to the three market indices, and the coefficients on the unexpected changes in monetary policy are very similar in magnitude to the case without interaction terms. Hence the asymmetry that affects the digital asset markets and the NASDAQ index, at least during the period examined, is the increase in the federal funds rate.

\section{Monetary Policy Announcements and Digital Asset Return Volatility}

In this part, we examine the effects of monetary policy announcements on the volatility of the 5-minute returns of digital assets  during the FOMC statement and Minutes releases and compare to non-release days.

\subsection{Price Data and FOMC Statement Releases} 
\label{Price Data and FOMC Statement Releases}

\begin{figure}[t!]
	\centering
	\caption{Volatility of Digital Asset Returns - All FOMC Statement Releases}
        \includegraphics[width = 8in]{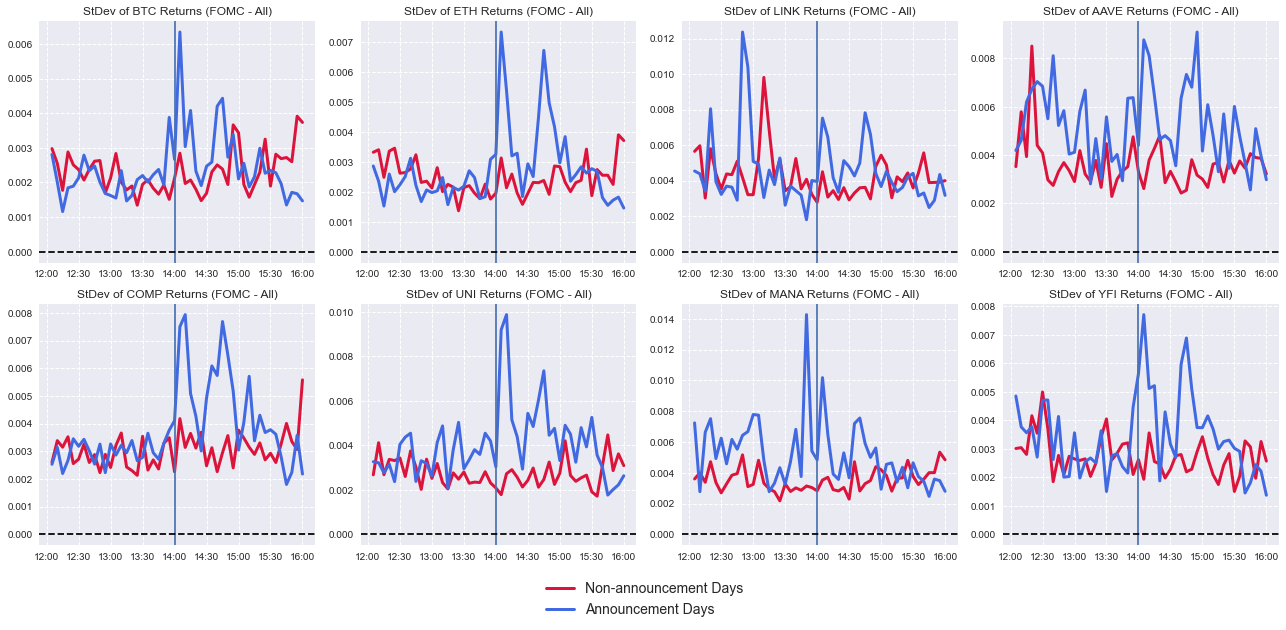}
	{\footnotesize \justify  \textit{Notes:} The above graphs depict the standard deviation of the 5-minute returns on various assets used as collateral on Aave, Compound, and Maker, on FOMC statement release days and non-release days for the whole period 01/01/2018-12/21/2022. \par}
\end{figure}

We use high-frequency asset price data, extracted from \textit{Coincap.io}, over 5-minute intervals to calculate the 5-minute returns on the assets used as collateral on the three DeFi protocols analyzed so far. In this part, we examine whether the volatility of digital asset prices is higher on the days the FOMC statements are released compared to non-announcement days. If the markets do not completely expect the contents of these releases, then investors are expected to rebalance their portfolios, and as a result, digital asset prices are expected to be more volatile at the time of a release. We follow \cite{rosa2013financial} and control for both intraday and day-of-the-week effects on asset prices since asset price volatility can vary with time. Specifically, we compute the standard deviation of 5-minute digital asset returns for the time interval 12:00-16:00, which starts two hours before any announcement and ends two hours after any announcement on event days. Then we compare it to the same measure computed at the same times on non-event days, which include the same weekday but one week before and one week after the announcement. Again, we examine the FOMC statements that were released between 01/01/2018 and 12/21/2022, under the condition that they contained a policy announcement about the federal funds rate or the central bank asset purchases.

\begin{figure}[t!]
	\centering
	\caption{Volatility of Digital Asset Returns - FOMC Statement Releases Before 12/15/2021}
	\includegraphics[width = 8in]{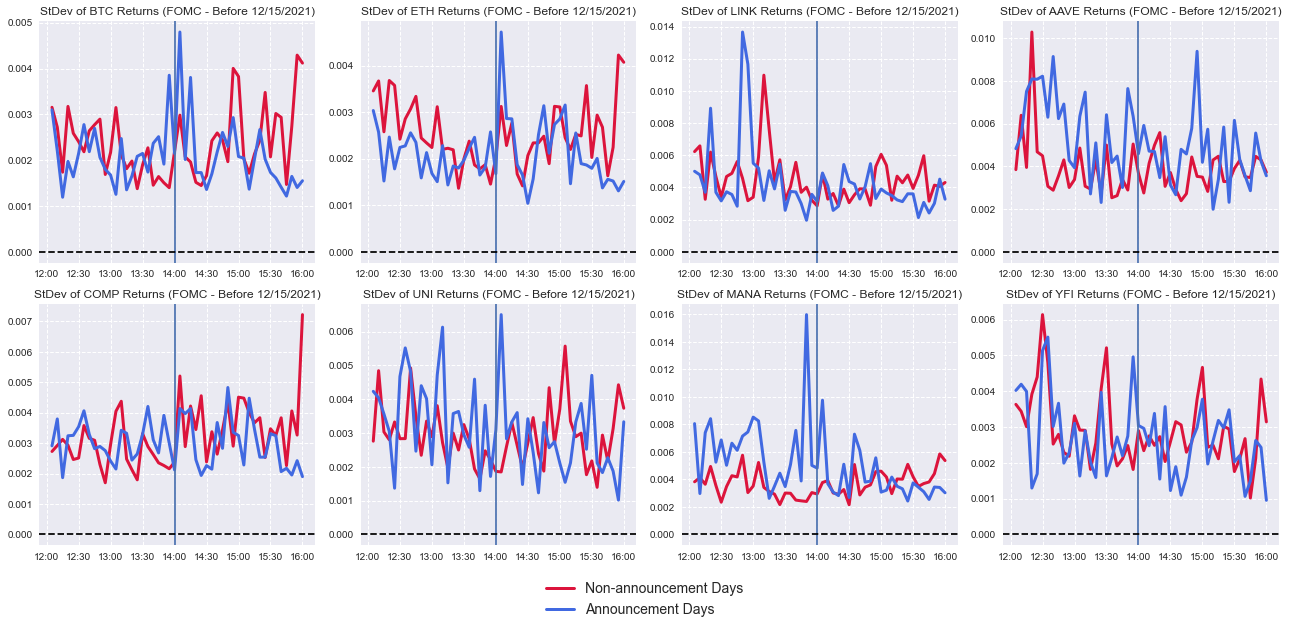}
	{\footnotesize \justify  \textit{Notes:} The above graphs depict the standard deviation of the 5-minute returns on various assets used as collateral on Aave, Compound, and Maker, on FOMC statement release days and non-release days for the period 01/01/2018-11/03/2021. \par}
\end{figure}

Figure 2 displays the results. The time of the release of the FOMC statement is depicted with the teal vertical line, which corresponds to the time of the announcement at 14:00. The graphs in Figure 2 show clearly that the volatility of digital asset returns increases significantly at the time of the FOMC statement release and up to one hour before and one hour after the release for some of the assets. Specifically, the volatility of asset prices increases significantly precisely at the time of the release and half an hour after the release, when the press conference takes place, and the chairman of the Federal Reserve explains the rationale of the monetary policy decisions publicly. An exception to the previous is the volatility of the MANA price, which peaks slightly before the FOMC statement release, but still increases at the time of the release. For all assets, the volatility of the returns is at least twice higher on announcement days. In addition, the spikes in the volatility of the altcoins are of more considerable magnitude than the spikes in the volatility of the BTC price, which indicates the more risky nature of these assets. 

We also investigate how the reaction of digital asset prices had changed after the FOMC announcement on 12/15/2021 when the Federal Reserve signaled a change in monetary policy that would lead to a more hawkish stance and hikes in the federal funds rate to fight the rising inflation. Figure 3 contains the volatility of the digital asset prices as measured for all the FOMC statement releases between 01/01/2018 and 11/03/2021. In contrast, Figure 4 contains the same measure for all the FOMC statement releases starting from 12/15/2021 until 12/21/2022.

\begin{figure}[t!]
	\centering
	\caption{Volatility of Digital Asset Returns - FOMC Statement Releases Since 12/15/2021}
	\includegraphics[width = 8in]{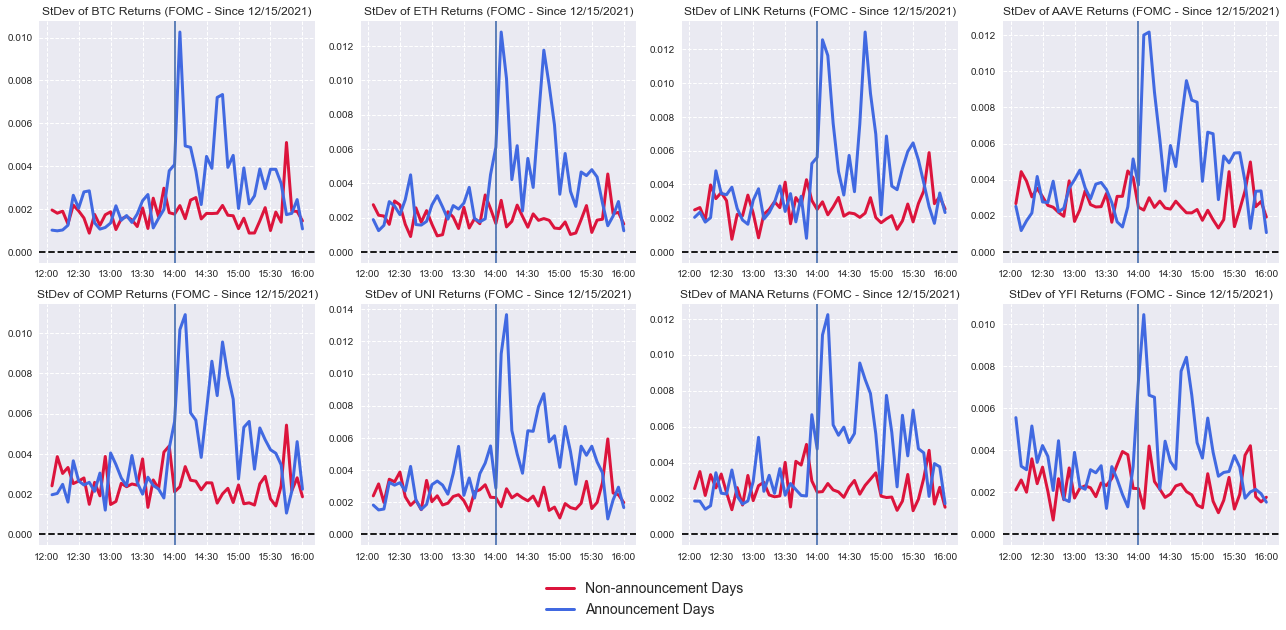}
	{\footnotesize \justify  \textit{Notes:} The above graphs depict the standard deviation of the 5-minute returns on various assets used as collateral on Aave, Compound, and Maker, on FOMC statement release days and non-release days for the period 12/15/2021-12/21/2022. \par}
\end{figure}

According to Figure 3, the volatility of digital asset prices had been affected by the FOMC statement releases in the past, but less significantly. Again, for most assets, except for LINK, there are spikes in the standard deviation at the time of the release. However, these spikes are smaller than when all the announcements were examined in the previous case. This implies that a significant portion of the volatility can be attributed to the more recent announcements after the change to a more hawkish monetary policy stance. Indeed, this is depicted in Figure 4, where the volatility increases in a more significant way for all the 5-minute intervals starting from the announcement time until two hours later. Again, the volatility of the altcoin returns is more substantial than the BTC returns. Of course, the previous results are not unexpected. Higher interest rates imply higher expected returns on government bonds. Given that the returns on bonds are considered to be relatively safe, investors sell their holdings of cryptocurrencies, starting from the riskier altcoins.

\subsection{Minutes Releases}

In this part, we repeat the exercise of the previous subsection and examine the Minutes releases. Again, we first focus on all the releases from 01/01/2018 to 12/21/2022, and then, we split the sample into two sub-samples before and after 12/15/2021. Figure 5 depicts the volatility of 5-minute returns for all the Minutes releases. Again, there are spikes in volatility at the time of the release on the release days, but now there are similar spikes for non-release days for most digital assets. The increase in volatility is not of the same magnitude as in the FOMC statement release, implying that the announcements of monetary policy decisions are more critical for digital asset prices than the release of the central bank Minutes.

\begin{figure}[t!]
	\centering
	\caption{Volatility of Digital Asset Returns - All Minutes Statement Releases}
	\includegraphics[width = 8in]{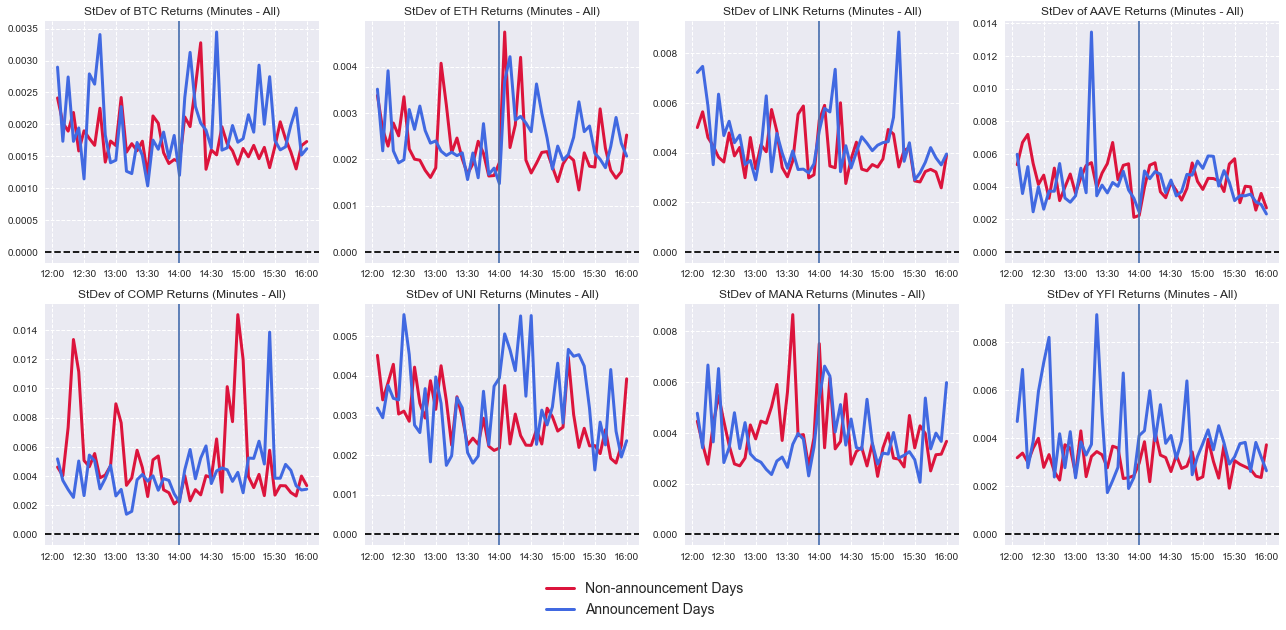}
	{\footnotesize \justify  \textit{Notes:} The above graphs depict the standard deviation of the 5-minute returns on various assets used as collateral on Aave, Compound, and Maker, on Minutes release days and non-release days for the whole period 01/01/2018-12/21/2022. \par}
\end{figure}

Next, we examine the Minutes releases for all the dates before 12/15/2021. The volatility graphs in Figure 6 remain virtually the same, as the differences are minor relative to Figure 5, implying that the release of the Minutes before the change in monetary policy stance did not significantly affect digital asset prices. On the other hand, the Minutes releases since 12/15/2021 seem to have more significant effects on volatility as shown in Figure 7, since for all assets considered, there are spikes in the standard deviation at the time of the release and up to one and a half hours after. One potential interpretation of this result is that the Minutes released after the threshold date contained details on the more hawkish views of the FOMC members, which led the investors holding more risky assets to offload their holdings as soon as they got the relevant information. Once again, the spikes in the volatility of altcoins are of a more considerable magnitude than the spikes in the volatility of BTC returns, reflecting the more risky nature of these assets.

\begin{figure}[t!]
	\centering
	\caption{Volatility of Digital Asset Returns - Minutes Releases Before 12/15/2021}
	\includegraphics[width = 8in]{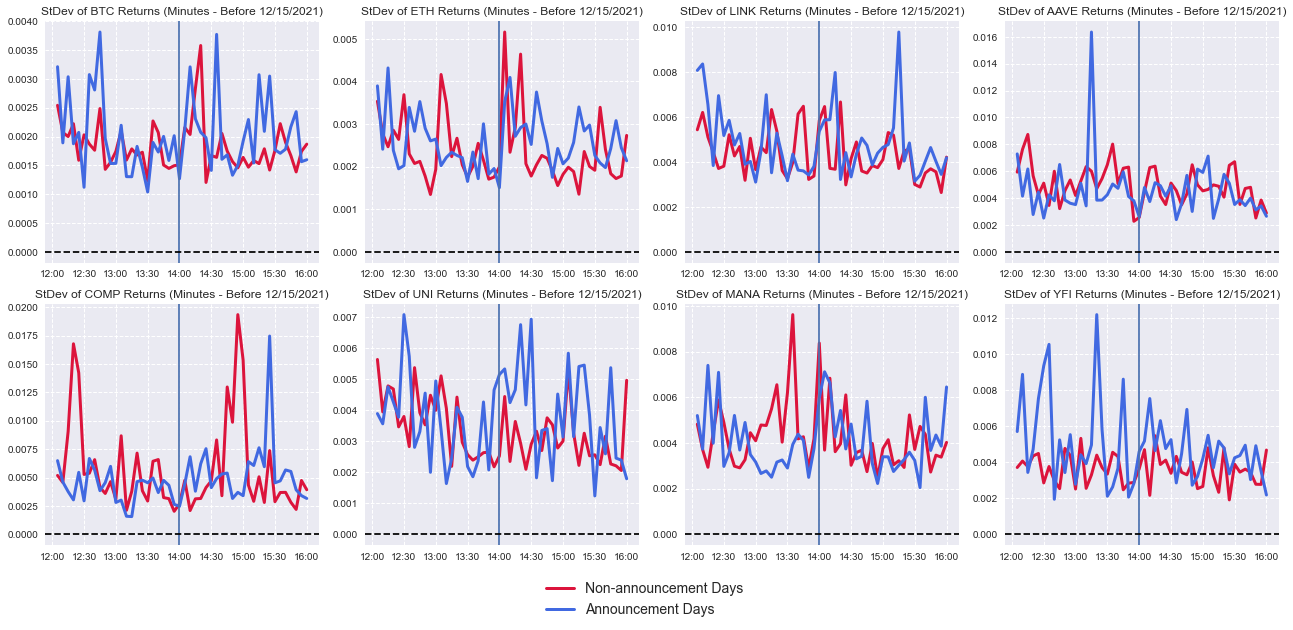}
	{\footnotesize \justify  \textit{Notes:} The above graphs depict the standard deviation of the 5-minute returns on various assets used as collateral on Aave, Compound, and Maker, on Minutes release days and non-release days for the period 01/01/2018-11/24/2021. \par}
\end{figure}

\begin{figure}[t!]
	\centering
	\caption{Volatility of Digital Asset Returns - Minutes Releases Since 12/15/2021}
	\includegraphics[width = 8in]{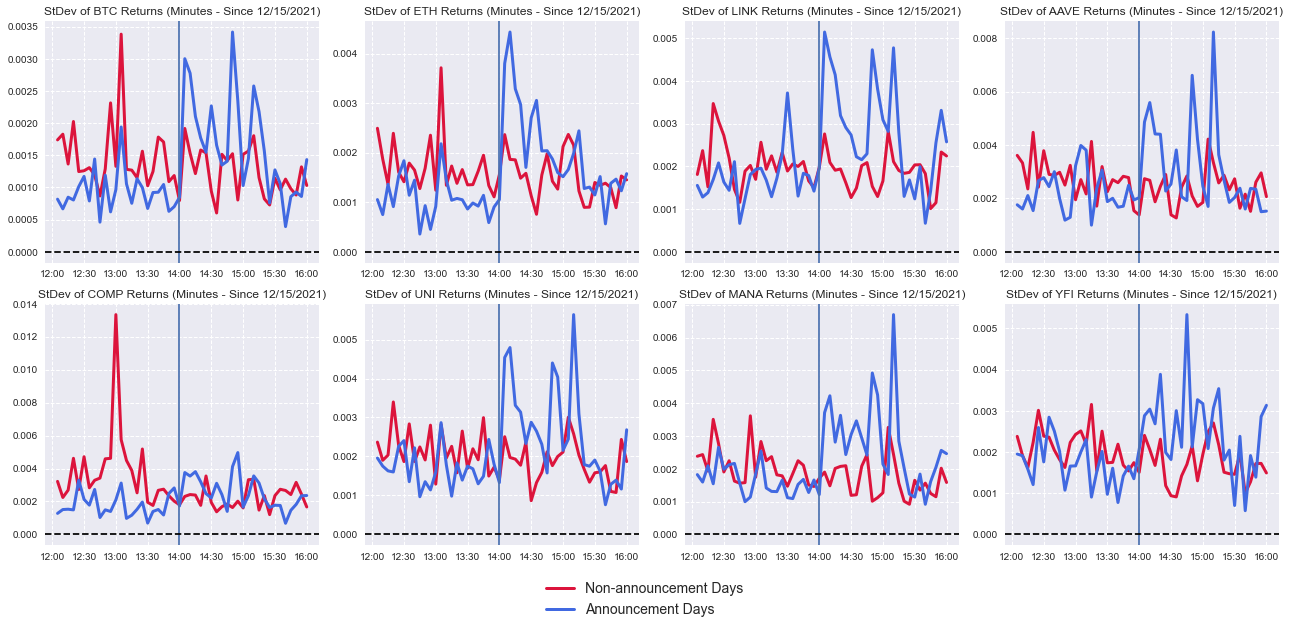}
	{\footnotesize \justify  \textit{Notes:} The above graphs depict the standard deviation of the 5-minute returns on various assets used as collateral on Aave, Compound, and Maker, on Minutes release days and non-release days for the period 12/15/2021-12/21/2022. \par}
\end{figure}

\section{Insights from Decentralized Finance}

Motivated by the results from the previous sections, we decided to examine how entities participating in Decentralized Finance react to changes in monetary policy.

\subsection{Data \& Methodology} \label{Data and Methodology}

Data used in this section originate from both the Ethereum blockchain and various complementary web resources. The reason for selecting Ethereum as the primary blockchain for this section is based on the fact that it hosts the majority of the DeFi platforms at the time of writing, as seen in platform \cite{DefiLlama}.

We set up a local Ethereum full archive node in our laboratory, in order to have unlimited access to all the transactions that took place since the first Ethereum block in July 2015, also called the Genesis block. This type of node is particularly useful when querying historical blockchain data, because it provides access to transaction data in its raw form, as it is stored in the blockchain, without the need for a third-party provider. However, this functionality comes at a cost, as it is resource-demanding; the current state of the node captures 2TB of storage and is continuously growing in size, as it is constantly syncing to the latest validated block.

By utilizing the node, we were able to search for transactions including specific actions that characterize the applications and protocols of Decentralized Finance. In this querying process for DeFi activities, we focused on the section of every transaction labeled log events. Smart contracts used in Ethereum blockchain have the ability to "emit" events during execution. The output of events is stored in transaction receipts under the Logs section. An event is indexed by its signature, a 256-bit hash, and the contract address emitting this event (\cite{GoEthereum.org}). Therefore, we can filter for the DeFi-specific events emitted from the studied protocols.

Throughout the extended list of available events from every protocol, we opted for ones that describe fundamental actions of Decentralized Finance. These are briefly summarized below.

\justify \textbf{Borrowing Funds:} This refers to the action of taking a loan of a specific asset available in the underlying protocol. Typically, borrowers are required to provide between 125\% to 150\% collateral of an asset \textit{x} to borrow 100\% of another asset \textit{y} (i.e., over-collateralization), as explained by \cite{qin_attacking_2021}.

\justify \textbf{Repaying a Loan:} After a user has borrowed funds from the protocol, it is expected of them to return these funds plus some interest, in order to receive their collateral back. Nonetheless, this is only possible if the value of their collateral has not dropped below a certain (liquidation) threshold, making this collateralized position prone to liquidation, which is explained below. 

\justify \textbf{Depositing Funds:} Users are able to deposit their assets into any DeFi protocol, instead of just keeping them into a crypto wallet, in order to earn interest on them.

\justify \textbf{Withdrawing Funds:} This relates to the action of removing funds from the protocol.

\justify \textbf{Liquidating:} In DeFi, similarly to Traditional Finance, a collateralized position can be liquidated, when a negative price fluctuation (a move below the liquidation threshold) of the debt collateral asset happens, as explained by \cite{qin_empirical_2021}. It is important to note that in permissionless blockchains, such as Ethereum, anyone can initiate a liquidation, repay the debt and claim the collateral.

\justify \textbf{Flash Loan:} Some DeFi protocols offer a unique type of loan, called flash loan, which enables the user to borrow funds without the need of a collateral. This kind of loan provides the borrower a considerable amount, only limited by the Total Value Locked in the pool at that time, as long as the loan plus some fees is repaid within the same blockchain transaction, as discussed by \cite{wang_towards_2021}.

\justify

We focused on 4 of the most prominent DeFi protocols, based on their TVL at the time of writing this work. These protocols are Aave, Compound, Maker and Liquity with \textit{6.82B} USD, \textit{2.78B} USD, \textit{7.21B} USD and \textit{600M} USD TVL respectively (\textit{\cite{DefiLlama}}). Moreover, the decision to focus on these protocols was also affected by the data availability each one of them provides. As events are not required in smart contracts, it is the choice of the protocol developers if they will add them and what portion of the data included in these actions will become publicly available. This fact greatly limits the research efforts in this area and underlines that the goal of permissionless blockchains for public data, like Ethereum, is still far from being achieved. 

Even with a local Ethereum node, one does not gain access to all the data we are interested in this study, such as DeFi APYs, crypto market indices, etc. For this reason, we used data from web resources and APIs that are referenced below.
The web API from \cite{Compound} protocol was used for historical borrow and supply rates of users for 6 basic assets, explained in section \ref{Federal Funds Rate, DeFi Borrowing Rates, Debt and TVL}. Furthermore, data from \textit{\cite{TheBlockResearch}} platform was used both for sections \ref{Federal Funds Rate, DeFi Borrowing Rates, Debt and TVL} and \ref{Events vs Announcements}. Lastly, we gathered input from free open-source APIs \textit{\cite{CoinCap.io}}, as mentioned in section \ref{Price Data and FOMC Statement Releases}, and \textit{\cite{Ethplorer.io}} for the purposes of crypto assets information and crypto markets capitalization respectively.

\begin{table}[t!]\centering
\def\sym#1{\ifmmode^{#1}\else\(^{#1}\)\fi}
\caption{Regression Results - Compound Borrowing Rates\label{cRATES}}
\begin{tabular*}{\hsize}{@{}l @{\extracolsep{\fill}} *{10}{c} @{}}
\toprule
                    &\multicolumn{1}{c}{(1)}&\multicolumn{1}{c}{(2)}&\multicolumn{1}{c}{(3)}&\multicolumn{1}{c}{(4)}&\multicolumn{1}{c}{(5)}&\multicolumn{1}{c}{(6)}&\multicolumn{1}{c}{(7)}&\multicolumn{1}{c}{(8)}&\multicolumn{1}{c}{(9)}&\multicolumn{1}{c}{(10)}\\
                    &\multicolumn{1}{c}{$i^{\text{cWBTC}}$}&\multicolumn{1}{c}{$i^{\text{cWBTC}}$}&\multicolumn{1}{c}{$i^{\text{cETH}}$}&\multicolumn{1}{c}{$i^{\text{cETH}}$}&\multicolumn{1}{c}{$i^{\text{cUSDC}}$}&\multicolumn{1}{c}{$i^{\text{cUSDC}}$}&\multicolumn{1}{c}{$i^{\text{cUSDT}}$}&\multicolumn{1}{c}{$i^{\text{cUSDT}}$}&\multicolumn{1}{c}{$i^{\text{cDAI}}$}&\multicolumn{1}{c}{$i^{\text{cDAI}}$}\\
\midrule
$\Delta i$                   &       -2.25\sym{*}         &                     &       0.30  &                     &       -6.55\sym{**} &           & 0.48       &   & 0.75   \\
                    &      (-2.59)         &                     &      (1.48)         &                     &      (-3.58)         &         & (0.17)       &    & (0.48)     \\
\addlinespace
$\Delta i^u$                  &                     &      15.09\sym{**}         &                     &      -0.024         &                     &      12.24 &       & -23.79       &  
    & -16.56\sym{*}\\
                    &                     &     (3.56)         &                     &     (-0.02)         &                     &     (1.06)   &        & (-1.63)   &   & (-2.15)   \\
\addlinespace
Intercept \hfill           &       3.373\sym{***}  &       2.815\sym{***} &       2.645\sym{***}  &       2.674\sym{***} &      7.221\sym{***}         &       6.326\sym{***}     & 4.752\sym{***}  & 5.345\sym{***}  & 3.596\sym{***}    & 4.048\sym{***}\\
                    &      (10.64)         &      (9.42)         &      (35.86)         &      (34.08)         &     (10.84)         &      (7.80)      & (4.51)     & (5.19)    & (6.29)    & (7.46)\\
\midrule
Adj. $R^2$         &          0.1688         &          0.2936         &          0.0407         &          -0.0370         &          0.2971         &          0.0046     & -0.0036    & 0.0555    & -0.0283    & 0.1144\\
\bottomrule
\multicolumn{7}{l}{\footnotesize \sym{*} \(p<0.05\), \sym{**} \(p<0.01\), \sym{***} \(p<0.001\)}\\
\end{tabular*}
\end{table}

\subsection{Federal Funds Rate, DeFi Borrowing Rates, Debt and TVL } \label{Federal Funds Rate, DeFi Borrowing Rates, Debt and TVL}

\begin{table}[t!]\centering
\def\sym#1{\ifmmode^{#1}\else\(^{#1}\)\fi}
\caption{Regression Results - Compound Debt and TVL\label{cTVL}}
\begin{tabular*}{0.85\textwidth}{@{}l @{\extracolsep{\fill}} *{4}{c} @{}}
\toprule
                    &\multicolumn{1}{c}{(1)}&\multicolumn{1}{c}{(2)}&\multicolumn{1}{c}{(3)}&\multicolumn{1}{c}{(4)}\\
                    &\multicolumn{1}{c}{ cDEBT}&\multicolumn{1}{c}{cDEBT}&\multicolumn{1}{c}{cTVL}&\multicolumn{1}{c}{cTVL}\\
\midrule
$\Delta i$                   &       2.27  &                     &       2.23\sym{*}  &                     \\
                    &      (1.97)         &                     &      (2.28)         &                     \\
\addlinespace
$\Delta i^u$                  &                     &      -12.53  &                     &      -11.12\sym{*}  \\
                    &                     &     (-2.01)         &                     &     (-2.06)         \\
\addlinespace
Intercept            &      19.869\sym{***}         &       20.339\sym{***}         &       21.040\sym{***}        &       21.475\sym{***}  \\
                    &      (47.5)         &      (47.03)         &      (59.06)         &      (57.33)         \\
\midrule
Adj. $R^2$         &          0.0903         &          0.0951         &          0.1263         &          0.1009         \\
\bottomrule
\multicolumn{5}{l}{\footnotesize \sym{*} \(p<0.05\), \sym{**} \(p<0.01\), \sym{***} \(p<0.001\)}\\
\end{tabular*}
\end{table}

In this part, we examine how the unexpected changes in the federal funds rate affect the interest rates on DeFi platforms, the debt outstanding, and the total value locked (TVL). We focus on the Compound protocol for which we have data since 2019, implying more degrees of freedom to our regressions.\footnote{Our data for the same variables on Aave and Maker start in late 2020, and this implies that there are very few FOMC announcements to take into account until 2022, which would lead to few degrees of freedom in any regression.} We extract data on borrowing interest rates on the following assets: BTC, ETH, USDC, USDT, and DAI, that is the two largest digital assets and three of the most important stablecoins. We also extract data from the website of \cite{TheBlockResearch} on debt outstanding and TVL in the Compound protocol denoted as cDEBT and cTVL, respectively. Our motivation is to examine whether an unexpected increase in the federal funds rate could be related to an increase in DeFi rates, for instance, due to fewer funds flowing into DeFi protocols as the risk-free rates are higher, which would push the DeFi interest rates higher due to less supply of funds, and also examine how the total debt outstanding the TVL in the protocol would respond to unexpected changes in the federal funds rate. Our methodology is the same as in section 3.1.

Table \ref{cRATES} contains the results of the event study regressions related to the borrowing rates in the Compound protocol.\footnote{We did the same exercise for the supply rates on Compound, but all the results were statistically insignificant.} According to Table 7, the only borrowing interest rate that is positively related to unanticipated changes in the federal funds rate in a statistically significant way is the borrowing rate on wrapped BTC. In that case, an unexpected increase by 1\% of the federal funds rate would lead to an increase in the borrowing rate on WBTC by 15.09\%. The adjusted $R^2$ on this regression is close to 29.4\%, which shows that the volatility in the federal funds rate can explain a significant part of the WBTC's borrowing rate volatility. 

The other borrowing rate, which is related in a statistically significant way to an unexpected change in the federal funds rate, is the borrowing rate on DAI. However, the relation is now negative. For the other three assets, the results are not statistically significant. Overall, the positive relationship between the interest rate on WBTC and the unexpected change in the federal funds rate reinforces our view that BTC behaves like a traditional asset with respect to other macro aggregates. 

Table \ref{cTVL} contains the regression results for the natural logarithm of the total debt outstanding and the natural logarithm of the total value locked in the Compound protocol. An unexpected increase by 1\% in the federal funds rate leads to a decrease in the total debt outstanding by around 12.5\% and a decrease in the TVL by around 11.1\%. The latter is a statistically significant move, whereas the former is not, but it is on the verge of being statistically significant at the 5\% level, with a $p$-value of 0.054. The adjusted $R^2$ in these two cases are equal to 9.51\% for the total debt outstanding and 10.09\% for the TVL. 

The intuition behind the previous results could be that as the federal funds rate increases unexpectedly, investors rebalance their portfolios and move funds outside of DeFi platforms, for instance, by favoring higher bond holdings, which leads to a decrease in the TVL in the platform. With lower resources available, activity is lower, and the debt is also lower. On the other hand, from Table 8, the relation between the TVL and the debt outstanding with the total change in the federal funds rate is positive, which is a less intuitive result.

\subsection{DeFi Activity \& FOMC Announcements} \label{Events vs Announcements}

Since results from the previous section suggest that changes in federal monetary policy could affect the activity in DeFi protocols, we were interested to examine how this translates to Ethereum as a whole, as well as to more specific DeFi actions in it. 

\begin{figure}[t]
    \centering
	\caption{Ethereum Daily Transactions over time and dates of FOMC Announcements}
    \includegraphics[width=\textwidth]{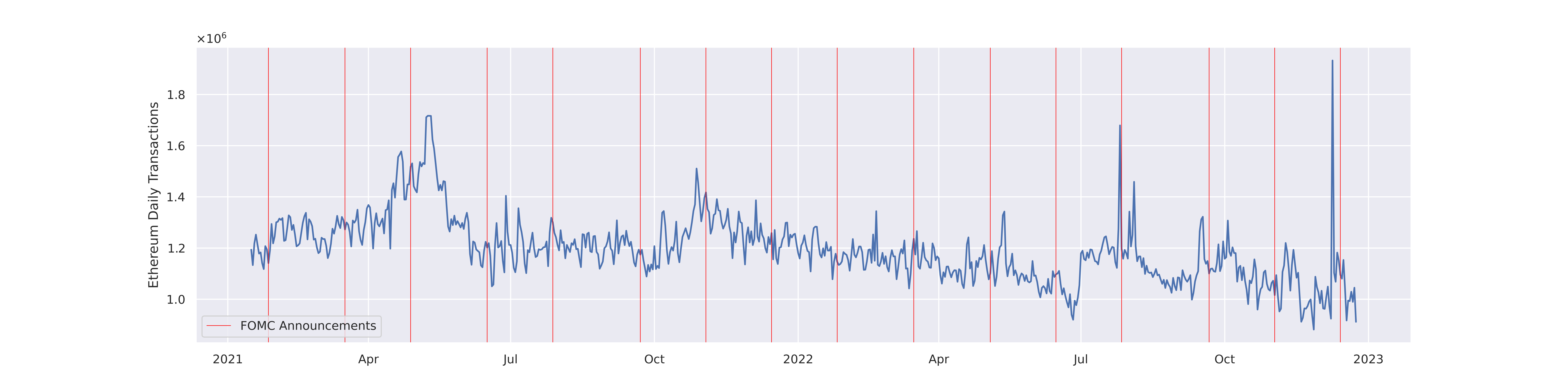}
    {\footnotesize \justify  \textit{Notes:} The figure above shows the number of daily transactions in Ethereum blockchain over time (blue line) for the time window of 01/01/2021-12/21/2022 and the corresponding FOMC announcements in this period. \par}
    \label{Ethereum Daily Transactions}
\end{figure}

Firstly, as one can see in figure \ref{Ethereum Daily Transactions}, we inspected the number of daily transactions in Ethereum (blue line) since 2021, compared to the dates of the monetary policy announcements (red vertical lines). The motivation behind this was to explore any noticeable trends in the data that could be explained by these announcements. Based on this figure, two out of the three highest values (1.6M and 1.9M transactions on July 26th and December 9th 2022 respectively), were only 1 and 5 days before the announcements on July 25th and December 14th 2022. This could mean that Ethereum users attempt to complete their transactions before any announcement regarding monetary policy is released. However, this daily volume of transaction data is "noisy", as it contains activity that is not necessarily related to finance. For example, Ethereum hosts a number of different decentralized applications, such as Gaming, Social Platforms, etc. Consequently, it was important to try and remove this "noise" from our data, by focusing on specific DeFi actions, as described in section \ref{Data and Methodology}.

\begin{figure}[t]
\centering
\caption{Daily Deposits \& Daily Withdrawals}
\begin{subfigure}{.45\textwidth}
  \centering
  \includegraphics[width=1\textwidth]{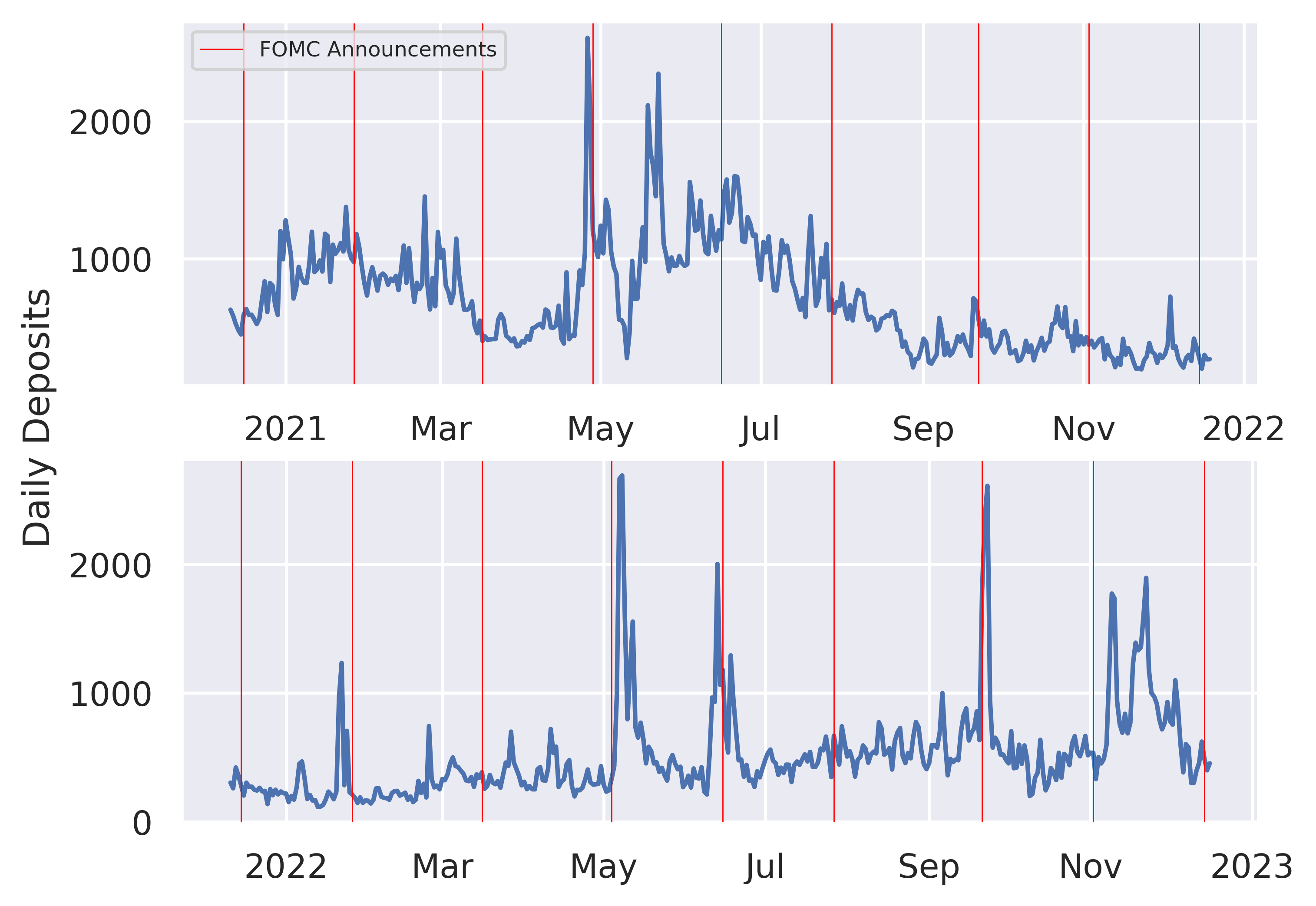}
  \caption{Daily Deposits}
  \label{Deposits_daily}
\end{subfigure}
\begin{subfigure}{.45\textwidth}
  \centering
  \includegraphics[width=1\textwidth]{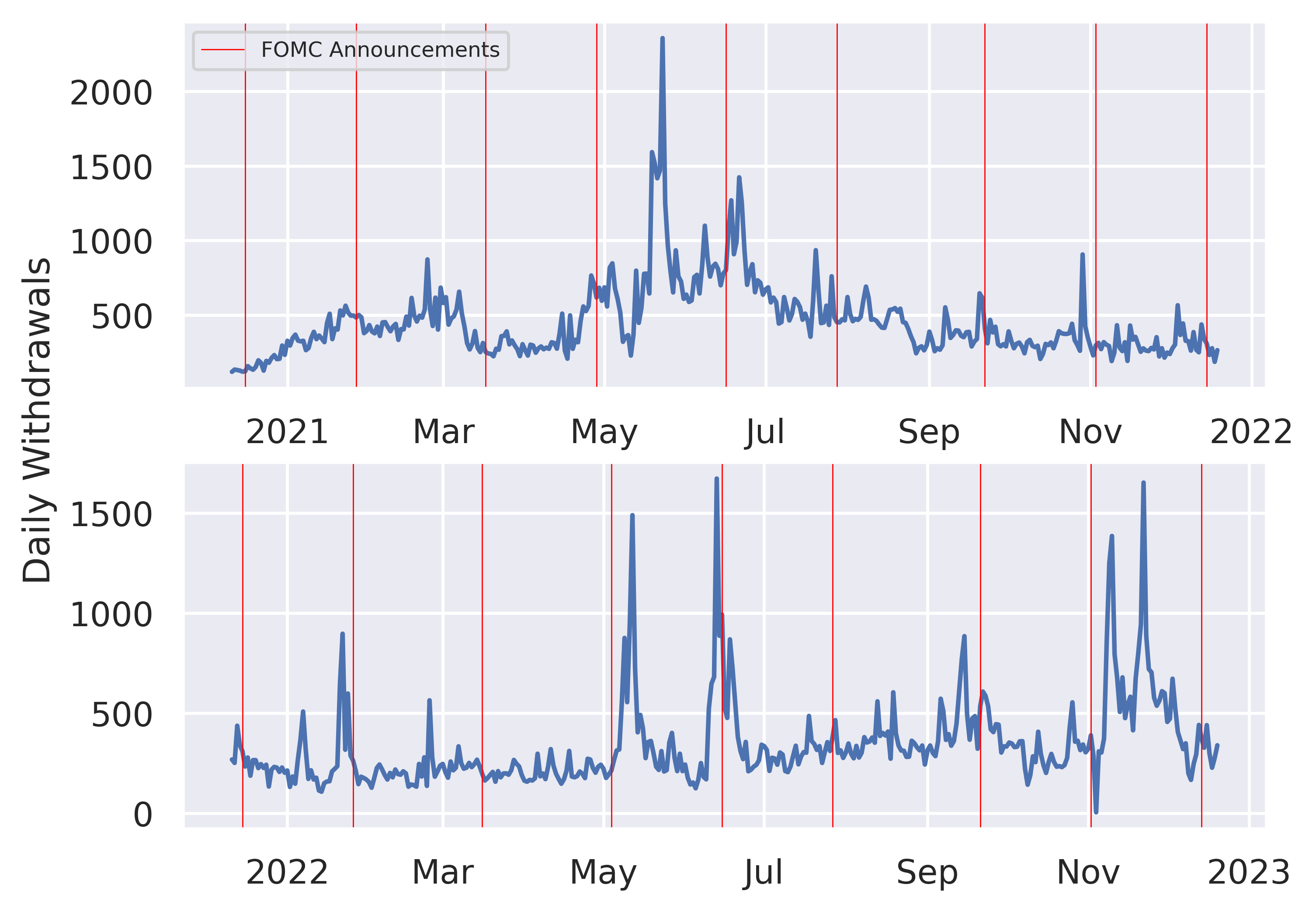}
  \caption{Daily Withdrawals}
  \label{Withdrawals_daily}
\end{subfigure}
{\footnotesize \justify  \textit{Notes:} This graph depicts the number of daily deposits (left subfigure) \& daily withdrawals (right subfigure) in Ethereum blockchain over time and the corresponding FOMC announcements compared between the two time windows of 01/01/2021 -- 12/15/2021 and 12/15/2021 -- 12/21/2022. \par}
\end{figure}

Starting with the deposits and withdrawals, we examined their number of daily events recorded in the protocols. The results are presented in figures \ref{Deposits_daily} and \ref{Withdrawals_daily}. In these figures we purposefully juxtapose these values between the two time windows of 01/01/2021 -- 12/15/2021 (upper subfigure)  and 12/15/2021 -- 12/21/2022 (lower subfigure), as they were also mentioned in section {section 5.1}. For deposits, it is distinct that while in the first time window there is no apparent trend close to FOMC announcements, this situation changes in the second one, as now at least 3 announcements (05/04/2022, 06/15/2022 and 09/21/2022) have high values close to them. For withdrawals, in a similar manner as deposits, we observe comparable behavior in activity between the two aforementioned time windows.

\begin{figure}[t]
\centering
\caption{Daily Loans \& Daily Loan Repayments}
\begin{subfigure}{.45\textwidth}
  \centering
  \includegraphics[width=1\textwidth]{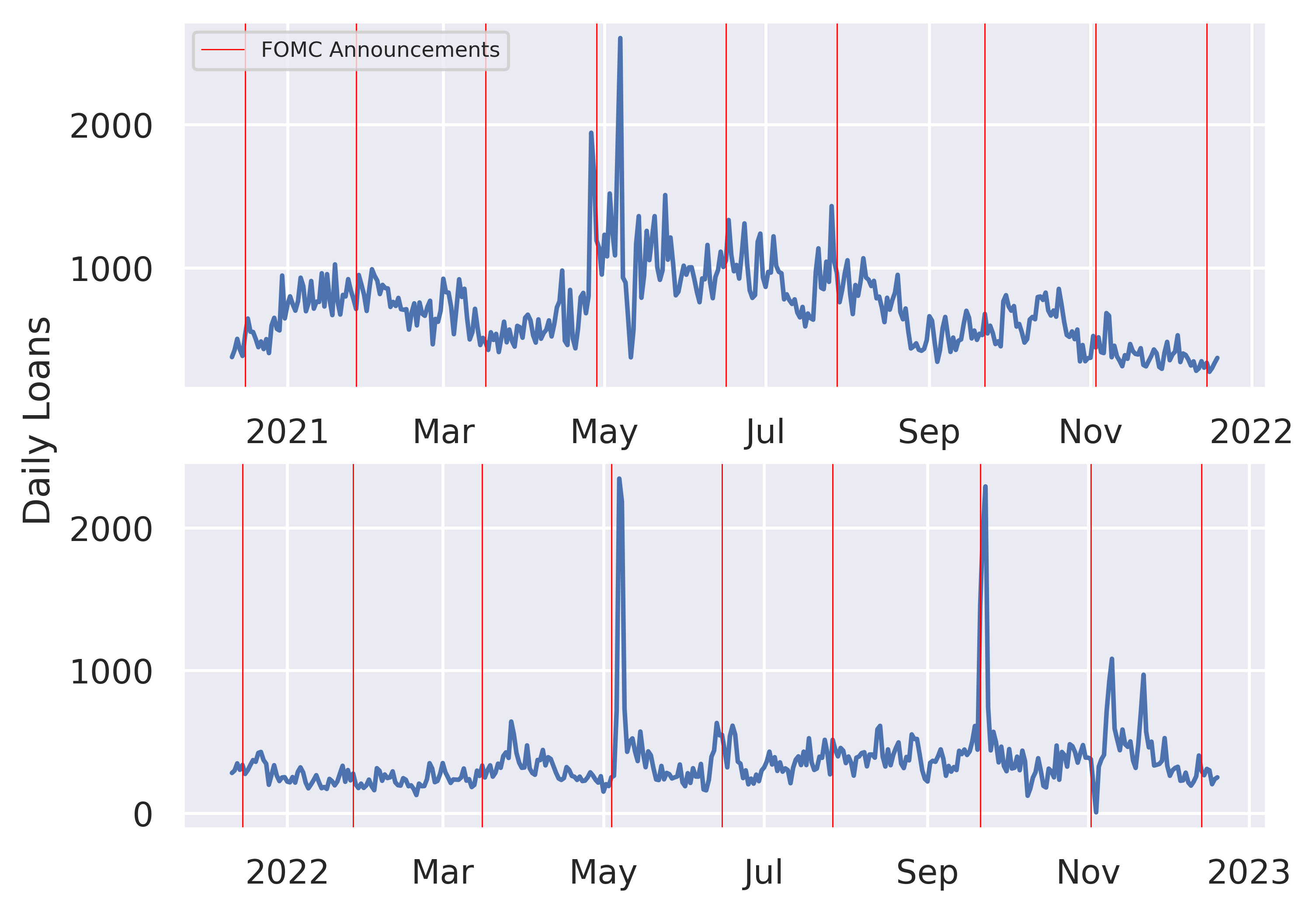}
  \caption{Daily Loans}
  \label{Borrows_daily}
\end{subfigure}
\begin{subfigure}{.45\textwidth}
  \centering
  \includegraphics[width=1\textwidth]{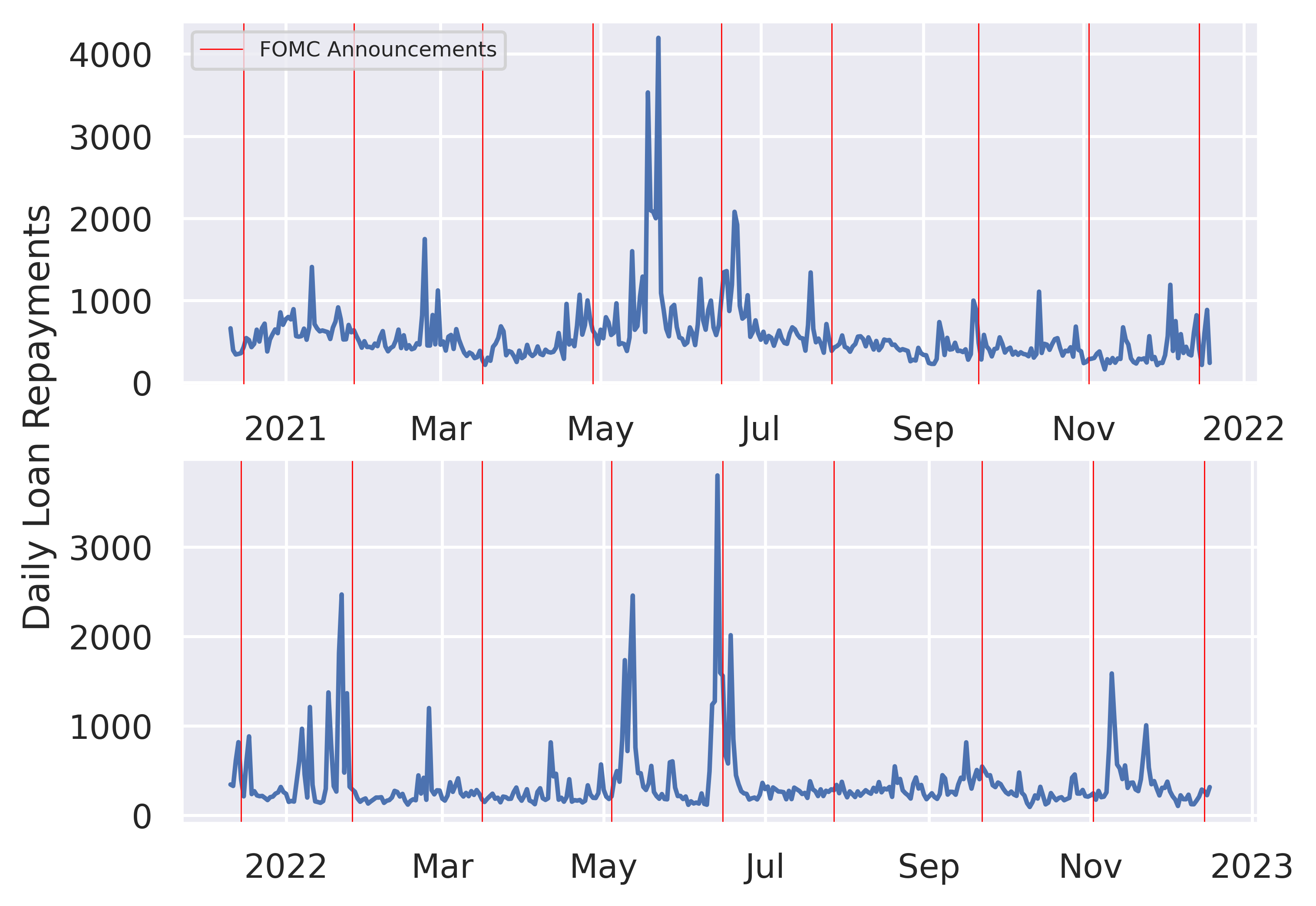}
  \caption{Daily Loan Repayments}
  \label{Repays_daily}
\end{subfigure}
{\footnotesize \justify  \textit{Notes:} This graph presents the number of daily loans (left subfigure) \& daily loan repayments (right subfigure) in Ethereum blockchain over time and the corresponding FOMC announcements compared between the two time windows of 01/01/2021 -- 12/15/2021 and 12/15/2021 -- 12/21/2022. \par}
\end{figure}

Moving on to loans and loan repayments, we provide the corresponding results in figures \ref{Borrows_daily} and \ref{Repays_daily}. For loan repayments, while there is some difference between the two time windows, there is no evident trend related to the FOMC announcements, except at 06/15/2022, where the daily number of loan repayments reached a peak just before the announcement. However, regarding the borrowing activity, we can inspect a definite contrast between these time windows, as the top-2 values are recorded just 3 and 2 days after the respective announcements on 05/04/2022 and 09/21/2022. Specifically, for the second one, the number of daily loans exhibits a relative increase of 226\% exactly on the date of the announcement. It is also important to notice that these 2 peaks are both observed after the respective dates, which implies that users wait for the announcements to be published and then engage in borrowing activities.

\begin{figure}[t]
\centering
\caption{Daily Liquidations \& Daily Flash Loans}
\begin{subfigure}{.45\textwidth}
  \centering
  \includegraphics[width=1\textwidth]{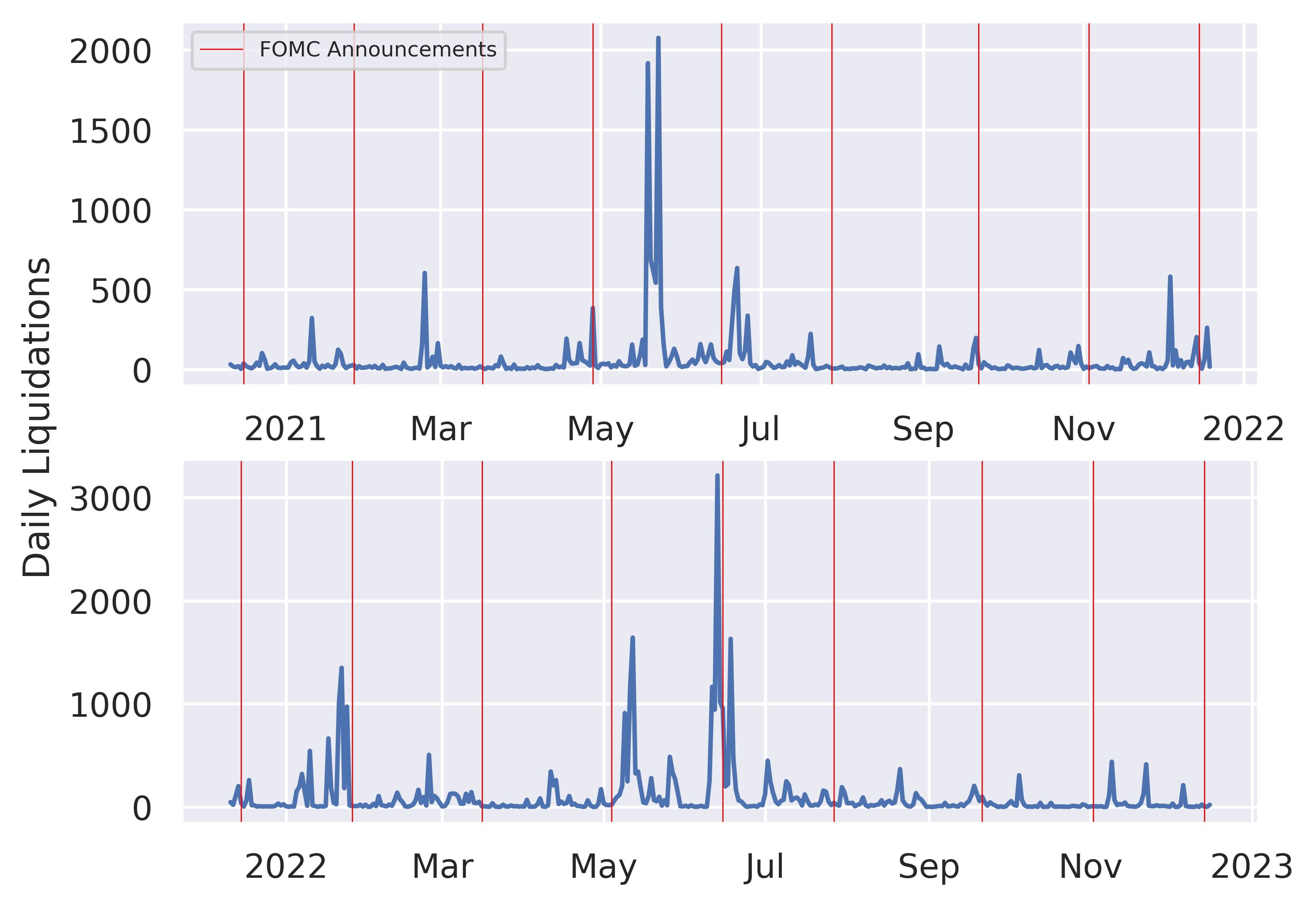}
  \caption{Daily Liquidations}
  \label{Liquidations_daily}
\end{subfigure}
\begin{subfigure}{.45\textwidth}
  \centering
  \includegraphics[width=1\textwidth]{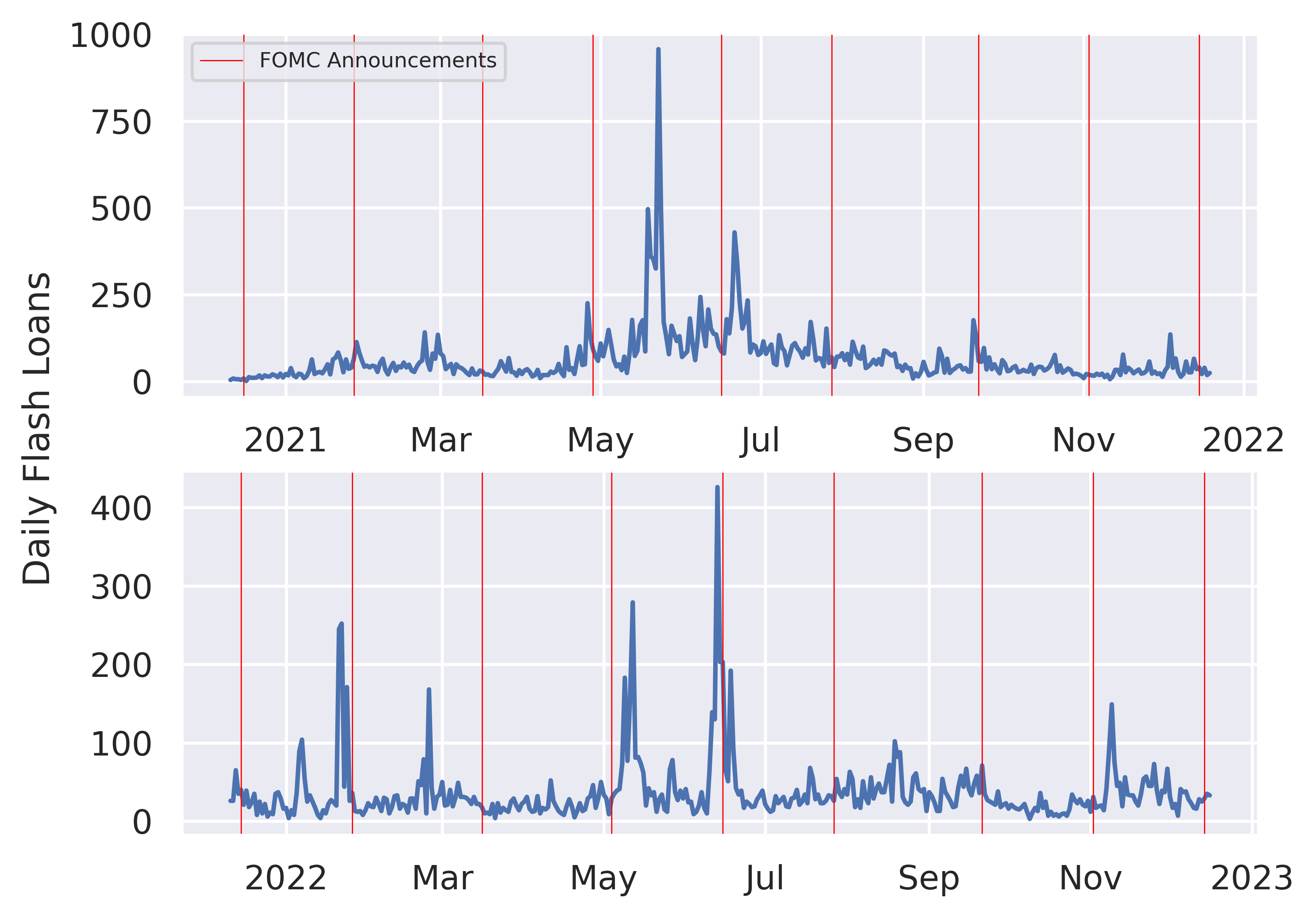}
  \caption{Daily Flash Loans}
  \label{Flashloans_daily}
\end{subfigure}
{\footnotesize \justify  \textit{Notes:} This graph shows the number of daily liquidations (left subfigure) \& daily flash loans (right subfigure) in Ethereum blockchain over time and the corresponding FOMC announcements compared between the two time windows of 01/01/2021 -- 12/15/2021 and 12/15/2021 -- 12/21/2022. \par}
\end{figure}

Before we continue with liquidations, it is important to provide some valuable insights from our data recorded for this action. As all of the studied protocols provide a Liquidation mechanism, we observed \textit{31,854}, \textit{8,797}, \textit{33,268} and \textit{849} successful liquidation events for Aave, Maker, Compound and Liquity respectively for the time window of 04/2019 -- 12/2022. We observed 398 unique assets used as collateral, with the most popular one being ETH among all the protocols, amounting for 45\% of all liquidated vaults. Regarding owners of vaults liquidated, we find \textit{18,582} unique users liquidated. On the contrary, if we assume that each unique Ethereum address represents one liquidator, we identify a total of \textit{1,749} unique liquidators, substantially less than that of the liquidated users. Interestingly enough, we discover that 10\% of liquidators are responsible for 85.4\% of the liquidations. This could mean that the whole liquidation mechanism in the DeFi scene has not yet been discovered by the wider audience, leaving the liquidation rewards to a small number of users. 

Lastly, the plots for liquidations and flash loans compared to FOMC announcements are presented in figures \ref{Liquidations_daily} and \ref{Flashloans_daily}. These two actions are similar to loan repayments discussed in the previous paragraph, meaning that, while they do show minor differences between the two time windows, they do not exhibit such strong trends regarding the FOMC announcements. Nevertheless, it is worthy of mention that these two events have great similarities in activity between them in the second time window, as can be seen in the respective bottom subfigures.

\section{Conclusion}

In this paper we focused on the effects that unexpected changes in monetary policy have on digital asset prices, and on how decentralized finance activity evolves around the FOMC announcements. The returns on BTC and ETH are negatively affected by surprises in monetary policy in a statistically significant way. The returns on the digital asset market indices are also negatively affected but the results are not statistically significant, although they are on the verge of being significant. We also show that the volatility of the 5-minute returns on digital assets increases significantly during the FOMC announcements, while the volatility of effect of the Minutes release is lower.

In the last part we show that some borrowing interest rates in the DeFi ecosystem are affected positively and in a statistically significant way. The debt outstanding and the TVL, on the other hand, are affected negatively. Finally, by utilizing a local Ethereum node,we investigated how the FOMC announcements influence the activity in 4 popular DeFi protocols. By plotting the daily number of basic DeFi actions over time, such as depositing and borrowing crypto assets, we demonstrated how FOMC announcements impact these values and set trends, particularly in the time period after December 2021.


\newpage
\bibliography{references}

\end{document}